\documentclass[aps,pra,twocolumn,groupedaddress,sort&compress,numbers]{revtex4-2}
\usepackage{graphicx}
\usepackage{subcaption}
\usepackage{dcolumn}
\usepackage{bm}
\usepackage[utf8]{inputenc}
\usepackage[T1]{fontenc}
\usepackage{array, float, tabularx, lipsum, amsmath, multirow} 
\usepackage{siunitx, xcolor}
\usepackage[version=4]{mhchem}
\graphicspath{{figs/}{figsgaoerb/}} 
\usepackage[colorlinks,linkcolor=blue,anchorcolor=blue,citecolor=blue]{hyperref}
\usepackage{svg}
\usepackage{ragged2e}
\usepackage{caption} 



\begin{document}

\title{Analyzing the performance of CV-MDI QKD under continuous-mode scenarios}


\author{Yanhao Sun$^{1}$}
\author{Ziyang Chen$^{2}$}
\email{chenziyang@pku.edu.cn}
\author{Xiangyu Wang$^{1}$}
\email{xywang@bupt.edu.cn}
\author{Song Yu$^{1}$}
\author{Hong Guo$^{2}$}

\affiliation{$^{1}$State Key Laboratory of Information Photonics and Optical Communications, Beijing University of Posts and Telecommunications, Beijing 100876, China. }
\affiliation{$^{2}$State Key Laboratory of Advanced Optical Communication Systems and Networks, School of Electronics, and Center for Quantum Information Technology, Peking University, Beijing 100871, China.}


\date{\ 24 January 2025}

\begin{abstract}
Continuous-variable measurement-device-independent quantum key distribution (CV-MDI QKD) can address vulnerabilities on the detection side of a QKD system. 
The core of this protocol involves continuous-variable Bell measurements performed by an untrusted third party. 
However, in high-speed systems, spectrum broadening causes Bell measurements to deviate from the ideal single-mode scenario, resulting in mode mismatches, reduced performance, and compromised security. 
Here, we introduce temporal modes (TMs) to analyze the performance of CV-MDI QKD under continuous-mode scenarios. 
The mismatch between Bob's transmitting mode and Bell measurement mode has a more significant effect on system performance compared to that on Alice's side. 
When the Bell receiver is close to Bob and the mismatch is set to just 5\%, the transmission distance drastically decreases from 87.96 km to 18.50 km. 
In comparison, the same mismatch for Alice reduces the distance to 86.83 km.
This greater degradation on Bob's side can be attributed to the asymmetry in the data modification step.
Furthermore, the mismatch in TM characteristics leads to a significant reduction in the secret key rate by 83\% when the transmission distance is set to 15 km, which severely limits the practical usability of the protocol over specific distances. 
These results indicate that in scenarios involving continuous-mode interference, such as large-scale MDI network setups, careful consideration of each user's TM characteristics is crucial. 
Rigorous pre-calibration of these modes is essential to ensure the system's reliability and efficiency.
\end{abstract}

\maketitle


\section{\label{I. Introduction}Introduction}
Quantum key distribution (QKD)~\cite{bennett2014quantum, 1ekert1991quantum, 2QKD_gisin2002quantum,3QKD_pirandola2020advances,4QKD_xu2020secure} can promise information-theoretic secure key distribution between legitimate 
communication parties (commonly referred to as Alice and Bob) when combined with 
the one-time pad encryption algorithm~\cite{5OTP_shannon1949communication}. 
Therefore, QKD can counter the potential threat posed by quantum computing to traditional cryptographic methods~\cite{6shor1994algorithms}.  
Continuous-variable (CV) QKD~\cite{7CVQKD_grosshans2002continuous,8CVQKD_weedbrook2004quantum} has garnered significant attention 
in recent years due to high compatibility with classical coherent optical communication components.
Although research on CV QKD has made remarkable progress in theory~\cite{9leverrier2013security,10leverrier2015composable,11leverrier2017security,12pirandola2017fundamental,ghorai2019asymptotic,lin2019asymptotic,upadhyaya2021dimension,denys2021explicit,lupo2022quantum,25chen2023continuous,zhang2023automatic,kanitschar2023finite,primaatmaja2024discrete}, 
experimentation (based on Gaussian modulation ~\cite{13jouguet2013experimental,qi2015generating,soh2015self,14huang2016long, 16hajomer2024long,williams2024field} and discrete modulation ~\cite{15wang2022sub,tian2023high, xu2023simultaneous, jaksch2024composable,hajomer2024experimental}),
post-processing~\cite{17post_wang2022continuous,18post_cao2023rate,19post_wang2023non}, 
and network applications~\cite{20network_wang2023experimental,21network_xu2023round,22network_qi2024experimental,23network_li2024experimental,bian2023high,hajomer2024continuous,kanitschar2024security},
new challenges have arisen with the advent of 
spectrum broadening~\cite{24specbroad_liu2022impact,25chen2023continuous}, 
which alters the traditional single-mode light field assumption.

In the security and performance analysis of CV QKD, 
the single-mode assumption is a fundamental and common premise. 
However, with the development of high-speed CV QKD systems, high-speed modulation introduces a nonuniform temporal waveform, 
causing the optical field to contain multiple frequency components, 
making the single-mode assumption no longer applicable.

Continuous-variable measurement-device-independent (CV-MDI)~\cite{26MDI_li2014continuous,27MDI_pirandola2015high,28MDI_lupo2018parameter,29MDI_tian2022experimental} protocol, 
a potential candidate for building future key distribution networks, 
is based on the ideal Bell measurement~\cite{30BellM_furusawa1998unconditional,31BellM_polkinghorne1999continuous}.
The Bell measurement involves the interference of two optical fields, which are traditionally assumed to be single-mode. 
However, with the advent of high-speed modulation~\cite{33highspeed_roumestan2024shaped,34highspeed_hajomer2024continuous} 
and non-ideal spectra, 
the issue of spectral broadening~\cite{35wright2017spectral} in the system has become prominent. 
The accuracy of Bell measurements is compromised in the continuous-mode framework. 
For a more realistic analysis of the CV-MDI protocol's performance, the practical spectral characteristics of devices, including light sources, detectors, and modulation, should be reconsidered, as these were previously ignored under the single-mode scenarios~\cite{25chen2023continuous}. 

Unlike the single-mode case, mode mismatches among the three parties involved in the protocol can significantly affect the outcomes of continuous-mode interference.
The single-mode Bell-measurement model cannot accurately reflect the inherent non-ideal characteristics of the devices. 
This discrepancy not only makes it difficult to align theoretical predictions with 
experimental results, but also complicates the performance and security analysis of the protocol.

Here, we propose using temporal modes (TMs)~\cite{36TM_blow1990continuum,37TM_brecht2015photon,38TM_fabre2020modes,39TM_raymer2020temporal,40TM_zhao2021propagation} 
to analyze CV-MDI QKD under continuous-mode scenarios. 
We establish an interference model for continuous-mode Bell measurements. 
Our model addresses how to quantitatively analyze the interference results of 
two broadband optical fields with temporal information under continuous-mode scenarios.
Furthermore, we apply this model to analyze the mode matching among the three parties. 
The mismatch involving Bob's TM has a more severe impact on the system's performance compared to that involving Alice's TM. 
Specifically, under the given conditions, a 5\% mismatch between Bob's transmitting mode and Bell measurement mode reduces the 
transmission distance from 87.96 km to 18.50 km, 
while the mismatch involving Alice reduces it only to 86.83 km. 
At 15 km, the former results in more than an 80\% reduction in the key rate compared to ideal conditions. 
This is because the data modification step in the CV-MDI QKD protocol is asymmetric, only Bob modifies his data (see Sec.~\ref{II. B} for details on the specific operations), while Alice's data remains unchanged.

This work presents an approach to analyzing the performance of the CV-MDI QKD protocol under practical conditions. 
The issues of non-ideal spectra and the spectrum broadening caused by high-speed modulation impose significant constraints on the protocol's practical usability. 
This work provides guidance for the design and optimization of future experiments,
rigorous pre-calibration of each user's TM characteristics and addressing these mismatches are essential steps to ensure the efficiency of the protocol in network configurations. 
\begin{figure*}[htb]
    \centering
    \includegraphics[width=\textwidth]{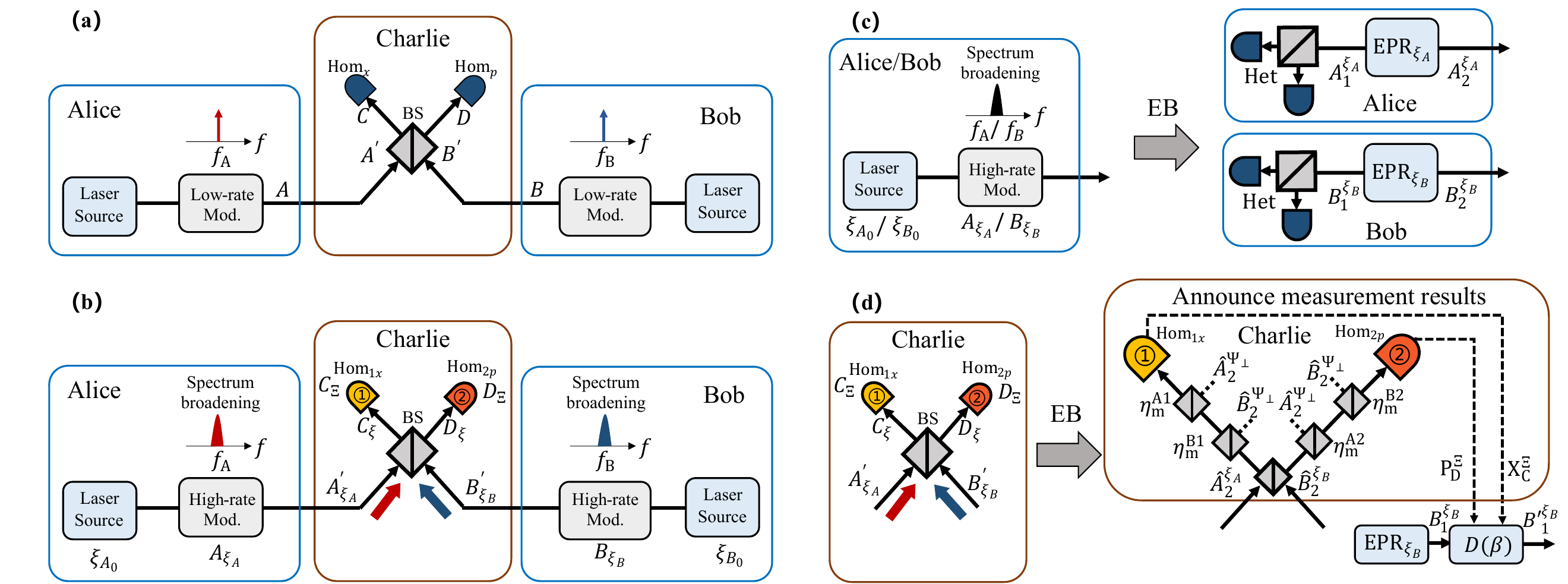}
    \captionsetup{justification=centering, format=plain} 
    \caption{
    \justifying 
    (a) PM model of CV-MDI under a single-mode scenario. 
    (b) PM model of CV-MDI under a continuous-mode scenario. 
    (c) The preparation of a continuous-mode Gaussian-modulated coherent state in the EB model 
    is equivalent to performing heterodyne detection on one mode of a continuous-mode TMSV.
    (d) Data correction in the PM model is equivalent to displacement operations in the 
    EB model. Mode-matching coefficients between the measured state's TM and the detector's TM is necessary under the continuous-mode scenario.
    }
    \label{PM in single/continuous, EB in continuous}
\end{figure*}

This paper is organized as follows: 
In Sec.~\ref{II. CONTINUOUS-MODE}, we analyze the CV-MDI QKD protocol under continuous-mode scenarios.
In Sec.~\ref{III. Numerical simulation}, we perform numerical simulations.
In Sec.~\ref{IV. RESULTS AND ANALYSIS}, we present the results of the secret key rate for different mode-matching coefficients and analyze the performance. 
Our conclusions are drawn in Sec.~\ref{V. CONCLUSION}.

\section{\label{II. CONTINUOUS-MODE}Continuous-Mode Analysis of CV-MDI QKD}
In this section, we first explain the changes in optical field modes in a practical 
high-speed CV-MDI QKD system. 
With the non-ideal spectrum and high-speed modulation, the optical field can be described 
in continuous-mode formalism~\cite{41TM_raymer1989temporal,36TM_blow1990continuum,39TM_raymer2020temporal}.
To provide a more practical analysis of the system's security and performance, we introduce TMs into the analysis.  
Then, we present the impact of spectrum broadening on both the entanglement-based (EB) scheme and the prepare-and-measure (PM) scheme of CV-MDI QKD. 
Finally, we provide the method for calculating the secret key rate of CV-MDI QKD under continuous-mode scenarios. 
When the effects of non-ideal experimental devices cannot be ignored, 
introducing TMs provides a framework for analyzing the performance of CV-MDI QKD under realistic experimental conditions.
Our analysis indicates that the assumption of single-mode light fields is a special case under this framework. 

\subsection{\label{II.A}The introduction of TMs}
The traditional security analysis is based on the assumption of single-mode optical fields, implying that the light beam has a single frequency. 
The ideal single-mode coherent state can be represented by the annihilation and creation operators of a single-mode field, $\hat{a}_i$ and $\hat{a}_i^{\dagger}$.

When a single-mode coherent state is modulated, new frequency components are introduced in 
the frequency domain, converting the single-mode coherent state into a multi-mode state. 
Furthermore, as the modulation speed increases, the introduced frequency components become 
more numerous and densely packed, causing the multi-mode state to approach a continuous-mode state with various frequency components. 
Under these circumstances, the single-mode field operator can no longer adequately describe 
a continuous-mode quantum state with some specific temporal waveforms. 
Therefore, a continuous-mode field operator is introduced to describe such states. 
By transforming the discrete-mode field operators~\cite{41TM_raymer1989temporal,36TM_blow1990continuum,39TM_raymer2020temporal,25chen2023continuous}, 
the continuous-mode annihilation and creation operators can be defined as 
$\hat{a}_i \rightarrow \sqrt{\Delta \omega} \hat{a}(\omega)$ and 
$\hat{a}_i^{\dagger} \rightarrow \sqrt{\Delta \omega} \hat{a}^{\dagger}(\omega)$,
where $\Delta \omega$ denotes the mode spacing.

To enable a more comprehensive analysis in the time domain, 
we take the creation operator as an example and perform an inverse Fourier transform on it, given by 
$\hat{a}^{\dagger}{ }(t)=({1}/{\sqrt{2 \pi}}) \int d \omega \hat{a}^{\dagger}(\omega) \exp (-i \omega t)$.
By defining a wave packet in the time domain as $\xi_i(t)$ 
(as an envelope $\xi_i^0(t)$ with a carrier $e^{-i \omega t}$), 
the photon-wavepacket creation operator~\cite{36TM_blow1990continuum,42loudon2000quantum} can be defined as 
\begin{equation}
\hat{A}_{\xi_i}^{\dagger}=\int d t \xi_i(t) \hat{a}^{\dagger}(t).
\end{equation}
The annihilation operator $\hat{A}_{\xi_i}$ follows a similar definition. 
Moreover, if $\xi_i(t)$ meets the orthonormalization, then $\hat{A}_{\xi_i}^{\dagger}$ and 
$\hat{A}_{\xi_i}$ are also known as the TM field operators~\cite{38TM_fabre2020modes,39TM_raymer2020temporal}. 
When $\hat{A}_{\xi_i}^{\dagger}$ acts on a vacuum state, it generates a coherent state with an envelope of $\xi_i(t)$.

Analyzing the protocol within the TM framework can illustrate the effects brought by the non-ideal spectrum and high-speed modulation. 

\subsection{\label{II. B}PM model with TMs}
Under the single-mode scenario, the PM scheme of the CV-MDI QKD protocol shown in Fig.~\ref{PM in single/continuous, EB in continuous}(a) is described as follows~\cite{26MDI_li2014continuous,27MDI_pirandola2015high}. 
First, Alice and Bob independently prepare Gaussian-modulated coherent states using their respective laser sources. 
Next, Alice and Bob send their coherent states to an untrusted third party, Charlie, through two separate channels. 
The two single modes ($A^{\prime}$ and $B^{\prime}$) received by Charlie interfere through a 50:50 beam splitter (BS), resulting in 
the output single modes $C$ and $D$. 
Charlie performs homodyne detections to measure the $x$ quadrature of $C$ and $p$ quadrature of $D$. 
After completing the measurements, Charlie announces the results 
$\left\{X_C, P_D\right\}$ and Bob modifies his data accordingly. 
Finally, Alice and Bob perform post-processing steps, such as parameter estimation, information reconciliation, and privacy amplification, to obtain the secret key.

However, the single-mode description does not capture time-domain information. 
Therefore, when using high-speed modulation (an inevitable trend for future development) 
in the CV-MDI QKD protocol, many issues cannot be ignored. 
The high-speed modulation introduces numerous emerged sidebands, leading to spectrum broadening, as shown in Fig.~\ref{PM in single/continuous, EB in continuous}(b). 
Under these conditions, Charlie's detection results will differ significantly from those expected under the single-mode assumption. 
Therefore, when taking into account high-speed modulation, spectral differences, and non-ideal detector responses, the single-mode assumption becomes invalid.
In contrast, a continuous-mode CV-MDI QKD model more accurately reflects the actual conditions than a single-mode model does.

The continuous-mode PM model shown in Fig.~\ref{PM in single/continuous, EB in continuous}(b), the effects of Alice's and Bob's non-ideal laser sources need to be considered. 
The quantum states generated by both parties are no longer single-mode but are continuous-mode coherent states. 
The main difference between single-mode and continuous-mode quantum states is that continuous-mode coherent states carry the time domain distribution characteristics of the optical field. 
The continuous-mode coherent states are sent to Charlie.

Within the TM framework, we can comprehensively describe continuous-mode coherent states. 
The quantum states transmitted through the channel can be expressed as~\cite{25chen2023continuous} 
$\left|x_A+i p_A\right\rangle_{\xi_A}$ and $\left|x_B+i p_B\right\rangle_{\xi_B}$, 
where $\xi_A$ and $\xi_B$ represent wave packets containing different temporal information. 
Charlie performs a continuous-mode Bell measurement on two states, 
which is crucial and also highlights issues neglected under the traditional single-mode scenario. 
Specifically, Charlie first interferes the two modes, $A_{\xi_A}^{\prime}$ and $B_{\xi_B}^{\prime}$, via a 50:50 beam splitter, 
with the interference outputs denoted as modes $C_{\xi}$ and $D_{\xi}$.
The two states with specific TMs from different senders cannot 
perfectly match the detectors' TMs at Charlie. 
This discrepancy affects the actual Bell measurement results and ultimately impacts the overall performance of a CV-MDI QKD system. 
On Charlie's side, both the $x$ quadrature of $C_{\xi}$ and the $p$ quadrature of $D_{\xi}$ are measured by homodyne detections.
Charlie announces the continuous-mode Bell measurement results $\left\{X_C^{\Xi}, P_D^{\Xi}\right\}$, 
which include not only the modulation information from both senders, 
but also the impact of the spectral characteristics of his own detector and the interfering signals. 
Thus, the entire continuous-mode protocol is inconsistent with the single-mode scenario.

In the CV-MDI QKD protocol, since Alice's and Bob's data are prepared independently, their initial data are completely uncorrelated. To establish correlations between Alice's and Bob's data,
after receiving the announced measurement results, Alice does not change her data, $X_A=x_A$, $P_A=p_A$.
While Bob modifies his data as $X_B=x_B+k_{\mathrm{m}}^B X_C^{\Xi}$, $P_B=p_B-k_{\mathrm{m}}^B P_D^{\Xi}$. 
The parameter $k_{\mathrm{m}}^B$ is a gain coefficient related to channel loss and the Bob's state's wave packet $\xi_B$. 
To achieve better performance, this modification needs to 
account for the impact of the broadened spectral characteristics on the continuous-mode Bell measurement. 
Alice and Bob use the modified data for parameter estimation, data reconciliation, and privacy amplification, 
ultimately obtaining the secret key.

The introduced TMs represent the time-domain distribution characteristics of the 
optical field, enabling the protocol to encompass more complex scenarios.

\subsection{\label{II. C}EB model with TMs}
The PM model is relatively easy to implement experimentally. 
By equating the preparation of a continuous-mode coherent state to performing 
heterodyne detection on one mode of a continuous-mode two-mode squeezed vacuum (TMSV) state ~\cite{grosshans2003virtual}
(as shown in Fig.~\ref{PM in single/continuous, EB in continuous}(c)), and simultaneously equating Bob's data correction operations to 
displacement operations on his mode~\cite{26MDI_li2014continuous} (as shown in Fig.~\ref{PM in single/continuous, EB in continuous}(d)), the EB model becomes equivalent to 
the PM model. This equivalence allows us to analyze the security of the protocol under a continuous-mode scenario. 
The EB model of CV-MDI QKD under a continuous-mode scenario is described as follows. 

\textbf{1. The preparation of continuous-mode quantum states.}
Alice and Bob each prepare $N$ continuous-mode TMSV states with variances 
$V_A$ and $V_B$, respectively. 
The continuous-mode TMSV states prepared by Alice and Bob can be represented as 
$\mathrm{EPR}_{\xi_A}$ and $\mathrm{EPR}_{\xi_B}$, 
where $\xi_A$ and $\xi_B$ represent the photon wave packets containing the 
temporal information of Alice's and Bob's light sources. 
Alice and Bob each keep one of their respective TMs, 
$A_1^{\xi_A}$ and $B_1^{\xi_B}$, while sending the other TMs 
($A_2^{\xi_A}$ and $B_2^{\xi_B}$) through two insecure channels to a completely untrusted third party, Charlie.

\textbf{2. Continuous-mode Bell Measurement.}
To better reflect practical experimental conditions, it is necessary to account for the limited bandwidth of the detectors and their non-ideal response functions. 
The impact of these non-ideal factors on detection efficiency is related to the mismatch between the measured state's TM and Charlie's detector's TM.
Its structure is illustrated in Fig.~\ref{PM in single/continuous, EB in continuous}(d). 

The wave packet form of the measured state's TM is $\xi$. 
The detection capability on Charlie's detectors is limited, leading to different 
detection efficiencies for different forms of wave packets. An ideal detector would detect all 
information from any wave packet, achieving 100\% detection efficiency at all times, 
which is a common assumption when describing the system using the single-mode model.

The introduction of TMs helps us identify Charlie's imperfect detection efficiency. 
The specific detection process can be equated to projecting the measured state's $\xi$-TM to 
the detector's $\Xi$-TM~\cite{25chen2023continuous}. 
The decomposition of the creation operator is as follows:
\begin{equation}
    \hat{A}_{\Xi}^{\dagger}=\sqrt{\eta_{\mathrm{m}}} \hat{A}_{\xi}^{\dagger}+\sqrt{1-\eta_{\mathrm{m}}} \hat{A}_{\Psi_{\perp}}^{\dagger},    
\end{equation}
where $\eta_{\text {m}}$ represents the mode-matching coefficient.
\begin{equation}
    \eta_{\mathrm{m}}=\left|\int d t \Xi^*(t) \xi(t)\right|^2.
\end{equation}

$\Xi^*(t)$ represents the conjugate of $\Xi(t)$, without considering any digital 
signal processing algorithms, $\Xi(t)$ can be expressed as 
\begin{equation}
    \Xi(t)=\frac{1}{\sigma_{\mathrm{cal}}} \xi_{\mathrm{LO}}(t) \exp \left(-i\left(\omega_{\mathrm{LO}} t\right)\right),
\end{equation}
where $\sigma_{\mathrm{cal}}=\sqrt{\int \mathrm{d} t\left|\xi_{\mathrm{LO}}(t)\right|^2}$ is the rescaled factor when calibrating output data by 
shot noise unit (SNU), $\xi_{\mathrm{LO}}(t)$ represents local oscillator's envelope, $\exp \left(-i\left(\omega_{\mathrm{LO}} t\right)\right)$ 
represents local oscillator's carrier, $\xi(t)$ represents the wave packet of the measured states.

In practical experiments, the impact of the mode-matching coefficient $\eta_{\mathrm{m}}$ is directly 
reflected in the first-order moments of the final data ($d_{\text {out }}$) and second-order moments of 
the final data ($\sigma^2$) ~\cite{25chen2023continuous}. $d_{\text {out }}=\sqrt{\eta_{\mathrm{m}}} d_{\text {in }}$, $d_{\mathrm{in}}$  denotes the mean value 
(first-order moment) of the input mode. $\sigma^2=\eta_{\mathrm{m}} V_{\mathrm{in}}+\left(1-\eta_{\mathrm{m}}\right)$ where $V_{\mathrm{in}}$ is the variance of the 
input mode.

Charlie performs a continuous-mode Bell measurement on the received states. 
Unlike the single-mode case, the continuous-mode scenario requires consideration of 
the mode-matching coefficients between the measured state's TM and the two detectors' TMs. 
More specifically, the four mode-matching coefficients that affect the detection results are shown in Table~\ref{TM Matching Coefficients}.

\begin{table}[h!]
    \centering
    \caption{TM Matching Coefficients}
    \begin{tabular}{c|c}
    \hline
    \textbf{Coefficient} & \textbf{Description} \\ \hline
    $\eta_{\mathrm{m}}^{\mathrm{A}_1}$ & Alice's $\xi_A$-TM vs. detector 1's $\Xi_1$-TM \\
    $\eta_{\mathrm{m}}^{\mathrm{A_2}}$ & Alice's $\xi_A$-TM vs. detector 2's $\Xi_2$-TM \\
    $\eta_{\mathrm{m}}^{\mathrm{B_1}}$ & Bob's $\xi_B$-TM vs. detector 1's $\Xi_1$-TM \\
    $\eta_{\mathrm{m}}^{\mathrm{B_2}}$ & Bob's $\xi_B$-TM vs. detector 2's $\Xi_2$-TM \\
    \hline
    \end{tabular}
    \label{TM Matching Coefficients}
\end{table}

Considering the mode-matching coefficients, Charlie receives the TMs $A_{\Xi}^{\prime}$ and 
$B_{\Xi}^{\prime}$. 
These two TMs then interfere at a 50:50 balanced beam splitter. 
Charlie then performs joint measurements on the output modes $C_{\Xi}$ and $D_{\Xi}$, 
obtaining the $x$ quadrature measurement result of mode $C_{\Xi}$ and the 
$p$ quadrature measurement result of mode $D_{\Xi}$. 
Afterward, Charlie announces the joint continuous-mode measurement results, 
$\left\{X_C^{\Xi}, P_D^{\Xi}\right\}$, to Alice and Bob through a public classical channel.

It can be seen that when each mode-matching coefficient $\eta_{\text{m}}$ equals 1, the protocol reduces to an ideal single-mode protocol.

\textbf{3. Displacement.}
After receiving Charlie's announced measurement results $\left\{X_C^{\Xi}, P_D^{\Xi}\right\}$, 
Bob performs a local displacement operation $D(\beta)$ on his mode $B_1^{\xi_B}$ to obtain $B_1^{\prime \xi_B}$. 
The displacement parameter $\beta$ is given by $\beta=g_{\text{m}}\left(x_C+p_D\right)$, 
where $g_{\text{m}}$ is the gain coefficient related to the overall channel parameters. 
It has been proven in previous works ~\cite{26MDI_li2014continuous} that when the parameter $g_{\mathrm{m}}$ in the 
EB model and the parameter $k_{\mathrm{m}}^B$ in the PM model satisfy the relationship 
$g_{\mathrm{m}}=k_{\mathrm{m}}^B \sqrt{\frac{V_B-1}{V_B+1}}$ (where $\left(V_B-1\right)$ represents the variance of Bob's modulation on the 
initial data $x_B$ and $p_B$), the joint probability distribution of all data 
$\left\{X_A, P_A, X_B, P_B, X_C^{\Xi}, P_D^{\Xi}\right\}$ is identical in the PM and EB models.
The optimal gain coefficient $g_{\text{m}}$ is also dependent on the mode-matching coefficients. 
In an actual experiment, $k_{\mathrm{m}}^B$ will be traversed to find an optimal value which 
makes the secret key rate the highest ~\cite{26MDI_li2014continuous}, $k_{\mathrm{m}}^B=\sqrt{\frac{2}{\eta_B \eta_{\mathrm{m}}^B}}$, $\eta_B$ represents Bob's channel 
transmittance, $\eta_B=10^{-\alpha L_{B C} / 10}$, where $\alpha$ represents the channel loss, and $L_{B C}$ denotes the 
distance between Bob and Charlie. $\eta_{\mathrm{m}}^B=\left|\int d t \Xi^*(t) \xi_B(t)\right|^2$ represents the mode-matching 
coefficient between Bob's TM and the detector's TM. 

\textbf{4. Postprocessing.}
Alice and Bob perform homodyne detection on their respective modes to obtain measurement data. 
Subsequently, they conduct parameter estimation and data post-processing 
(including data reconciliation and privacy amplification), ultimately obtaining the secret key.

In summary, Table~\ref{Comparison between Single-Mode and Continuous-Mode Scenarios} presents the differences between the single-mode and continuous-mode scenarios.

\begin{table*}[htbp]%
    \caption{\label{Comparison between Single-Mode and Continuous-Mode Scenarios}%
    Comparison between Single-Mode and Continuous-Mode Scenarios
    }
    \centering 
    \begin{ruledtabular} 
    \begin{tabular}{ccc} 
    \textnormal{} & \textnormal{Single-mode scenarios} & \textnormal{Continuous-mode scenarios} \\ 
    \colrule 
    \textnormal{State preparation in PM} & \textnormal{Single-mode coherent state} & \textnormal{Continuous-mode coherent state} \\ 
    \textnormal{State measurement in PM} & \textnormal{Ideal single-mode Bell measurement} & \textnormal{Continuous-mode Bell measurement} \\ 
    \textnormal{Data modification in PM} & \textnormal{Considering channel attenuation only} & \textnormal{Channel attenuation \& spectral mode mismatch} \\ 
    \textnormal{State preparation in EB} & \textnormal{Single-mode TMSV} & \textnormal{Continuous-mode TMSV} \\ 
    \textnormal{Bell measurement in EB} & \textnormal{Perfect mode matching} & \textnormal{4 mode-matching coefficients} \\ 
    \textnormal{Displacement in EB} & \textnormal{$g$} & \textnormal{$g_{\text{m}}$} \\ 
    \end{tabular}
    \end{ruledtabular}
\end{table*}

\section{\label{III. Numerical simulation}Numerical simulation}
\begin{figure}[htbp]
    \includegraphics[width=0.48\textwidth]{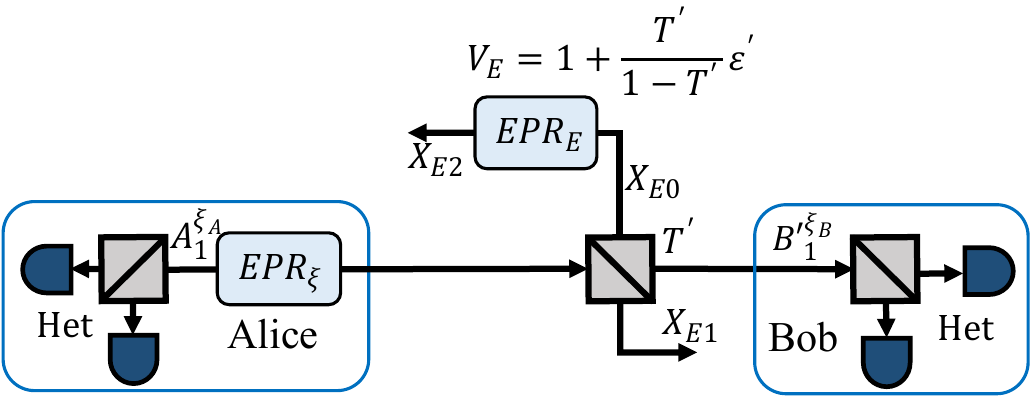} 
    \caption{
    \justifying 
    \label{equivalent one-way} Equivalent one-way model of the EB scheme.
    }
\end{figure}

After establishing the equivalence between the PM and EB models of the CV-MDI 
protocol, the security analysis can be conducted within the EB framework. 
The EB scheme of MDI QKD can be seen as a one-way protocol using entanglement swapping as 
an untrusted quantum relay ~\cite{furusawa1998unconditional,polkinghorne1999continuous,26MDI_li2014continuous}.
We assume that the eavesdropper, Eve, performs independent eavesdropping operations on the two channels separately, which accurately describes the actual system~\cite{26MDI_li2014continuous}. 
Furthermore, if both the preparation of Bob's EPR state and the displacement operations are assumed to be manipulated by Eve, 
the CV-MDI QKD protocol could be seen as the one-way CV-QKD protocol using coherent states and heterodyne detection. 
Its EB model is illustrated in Fig.~\ref{equivalent one-way}.
\noindent

$T^{\prime}$ represents the equivalent channel transmittance in the one-way model, 
and $\varepsilon^{\prime}$ represents the equivalent excess noise in the one-way model.

In this paper, we consider the secret key rate against single-mode attacks in the asymptotic limit. 
We use the reverse reconciliation method to calculate the secret key rate. The formula is given by~\cite{43Eq4_cover1999elements,devetak2005distillation}:
\begin{equation}
    K=\beta_R I(A: B)-\chi(B: E),
\end{equation}
where $\beta_R$ is the reconciliation efficiency, $I(A: B)$ is the mutual information between Alice and Bob, and $\chi(B: E)$ is the Holevo bound of the mutual information between Bob and Eve.
In the experiment, Alice and Bob obtain the covariance matrix 
$\gamma_{A_1^{\xi_A} B_1^{\prime \xi_B}}$ through the parameter estimation step, then 
calculate $I(A: B)$ and $\chi(B: E)$.

The quadratures' relations are shown in Appendix~\ref{Appendix A. quadratures}. 
The covariance matrix $\gamma_{A_1^{\xi_A} B_1^{\prime \xi_B}}$ of the equivalent 
one-way protocol is 
\begin{equation}
    \gamma_{A_1^{\xi_A} B_1^{\prime \xi_B}}=\left(\begin{array}{cc}A_1 & C \\ C^{T} & B_1^{\prime}\end{array}\right),
\label{covariance matrix}
\end{equation}
where $A_1=V_A I_2$ and $I_2$ represents \(2\times2\) identity matrix.
$C=\mathrm{diag}(a, b)$, 
$a=\sqrt{\eta_{\mathrm{m}}^{\mathrm{A} 1} T_x \left( V_A^2 - 1 \right)}$, and
$b=-\sqrt{\eta_{\mathrm{m}}^{\mathrm{A} 2} T_p \left( V_A^2 - 1 \right)}$.
$C^{T}$ represents the transpose of $C$.
$B_1^{\prime}=\mathrm{diag}(c, d)$,
$c=T_x\left(V_A-1\right)+1+T_x \varepsilon_x^{\prime}$, 
$d=T_p\left(V_A-1\right)+1+T_p \varepsilon_p^{\prime}$.

Channel parameters transmittance and excess noise on Alice's side 
(Bob's side) are $\eta_A$ ($\eta_B$) and $\varepsilon_A$ ($\varepsilon_B$). 
Assuming channel loss is $\alpha=0.2$ dB/km, $\eta_A=10^{-\alpha L_{A C} / 10}$ and $\eta_B=10^{-\alpha L_{B C} / 10}$. 

$T_x=\eta_{\mathrm{m}}^{\mathrm{A}_1}(\eta_A/{2}) (g_{\mathrm{m}}^x){ }^2$, 
$T_p=\eta_{\mathrm{m}}^{\mathrm{A}_2}(\eta_A/{2})(g_{\mathrm{m}}^p){ }^2$,
both $g_{\mathrm{m}}^x$ and $g_{\mathrm{m}}^p$ are displacement gain coefficients. 

\begin{align}
    &\varepsilon_x^{\prime} = 1 + \frac{1}{\eta_A} \left[\eta_B \left(\chi_B + 1 - 2 \eta_{\mathrm{m}}^{\mathrm{B} 1}\right) + \eta_A \chi_A \right] \notag \\
    &\quad + \frac{1}{\eta_A} \left(\frac{\sqrt{2}}{g_{\mathrm{m}}^x} \sqrt{V_B - 1} - \sqrt{\eta_{\mathrm{m}}^{\mathrm{B} 1}} \sqrt{\eta_B} \sqrt{\eta_B + 1}\right)^2
\end{align}
and
\begin{align}
    &\varepsilon_p^{\prime} = 1 + \frac{1}{\eta_A} \left[\eta_B \left(\chi_B + 1 - 2 \eta_{\mathrm{m}}^{\mathrm{B} 2}\right) + \eta_A \chi_A \right] \notag \\
    & \quad + \frac{1}{\eta_A} \left(\frac{\sqrt{2}}{g_{\mathrm{m}}^p} \sqrt{V_B - 1} - \sqrt{\eta_{\mathrm{m}}^{\mathrm{B} 2}} \sqrt{\eta_B} \sqrt{\eta_B + 1}\right)^2
\end{align}
are equivalent excess noise, 
$\chi_A=({1-\eta_A})/{\eta_A}+\varepsilon_A$ and $\chi_B=({1-\eta_B})/{\eta_B}+\varepsilon_B$.
   
In order to minimize the equivalent excess noise, we choose 
$g_{\mathrm{m}}^x = \sqrt{[2(V_B - 1)]/[\eta_{\mathrm{m}}^{\mathrm{B} 1}\eta_B(V_B + 1)]}$ and
$g_{\mathrm{m}}^p = \sqrt{[2(V_B - 1)]/[\eta_{\mathrm{m}}^{\mathrm{B} 2}\eta_B(V_B + 1)]}$;
thus we have: 
\begin{equation}
    \varepsilon_x^{\prime}=\varepsilon_A+\frac{1}{\eta_A}\left[\eta_B\left(\varepsilon_B-2 \eta_{\mathrm{m}}^{\mathrm{B} 1}\right)+2\right]    
\end{equation}
and
\begin{equation}
    \varepsilon_p^{\prime}=\varepsilon_A+\frac{1}{\eta_A}\left[\eta_B\left(\varepsilon_B-2 \eta_{\mathrm{m}}^{\mathrm{B} 2}\right)+2\right].         
\end{equation}

Unlike single-mode scenarios, to minimize the equivalent excess noise, 
Bob should consider the mode-matching coefficients between his TM and the detectors' TMs 
when performing the displacement operation on the retained mode. 

As discussed above, the four mode-matching coefficients between the 
senders and the detectors affect the protocol's performance. 
To clarify the impact of these mode-matching coefficients on the system's performance, 
we make reasonable assumptions for different scenarios to simplify the parameters, 
as outlined below:

At Charlie's side, detector 1 measures the $x$ quadrature, detector 2 measures the $p$ quadrature. 
We assume that both detectors have the same bandwidth, response function, and sampling rate. 
Since the uncertainty of a coherent state in the $x$ and $p$ quadratures is identical, 
swapping the mode-matching coefficients between the same quantum state and the 
two detectors in the simulation yields the same results. 
Conversely, an excessive number of parameters would increase the complexity of the analysis. 
Therefore, under our assumptions:
$\eta_{\mathrm{m}}^{\mathrm{A} 1}=\eta_{\mathrm{m}}^{\mathrm{A} 2}=\eta_{\mathrm{m}}^{\mathrm{A}}$;
$\eta_{\mathrm{m}}^{\mathrm{B} 1}=\eta_{\mathrm{m}}^{\mathrm{B} 2}=\eta_{\mathrm{m}}^{\mathrm{B}}$.

$\eta_{\mathrm{m}}^{\mathrm{A}}$ represents the mode-matching coefficient between 
Alice's $\xi_A$-TM and Charlie's $\Xi$-TM,
$\eta_{\mathrm{m}}^{\mathrm{B}}$ represents the mode-matching coefficient between 
Bob's $\xi_B$-TM and Charlie's $\Xi$-TM.
The simulation results and the detailed analyses are given in the next section.

\section{\label{IV. RESULTS AND ANALYSIS}RESULTS AND ANALYSIS}
\subsection{\label{IV. A}Impact of mode-matching coefficients on maximum transmission distance}
To analyze the impact of mode-matching coefficients on the maximum transmission distance, 
we consider the following scenarios: 
when $\eta_{\mathrm{m}}^{\mathrm{A}}=\eta_{\mathrm{m}}^{\mathrm{B}}=1$, it corresponds 
to the ideal single-mode case; when the mode-matching coefficients are less than 1, 
it indicates mode mismatch due to the effects of non-ideal laser sources and detectors. 
Mode mismatch reduces detection efficiency, thereby affecting the performance of the CV-MDI QKD protocol. 
This impact is difficult to notice under the traditional single-mode scenario.

We set four parameters, 
$\eta_{\mathrm{m}}^{\mathrm{A}}=\eta_{\mathrm{m}}^{\mathrm{B}}=1$; 
$\eta_{\mathrm{m}}^{\mathrm{A}}=0.95$, $\eta_{\mathrm{m}}^{\mathrm{B}}=1$; 
$\eta_{\mathrm{m}}^{\mathrm{A}}=1$, $\eta_{\mathrm{m}}^{\mathrm{B}}=0.95$; 
$\eta_{\mathrm{m}}^{\mathrm{A}}=\eta_{\mathrm{m}}^{\mathrm{B}}=0.95$, 
to analyze the impact of different degrees of mode matching on system performance, 
and considered three different configurations of the CV-MDI QKD protocol.
The other parameters are set as follows: 
$\beta_R=1$, $V_A=V_B=40$, $\varepsilon_A=\varepsilon_B=0.002$.

\textbf{Case 1:} Symmetric configuration. 
Figure.~\ref{C at mid, C at A}(a) illustrates the impact of mode-matching coefficients on the secret key rate. 
In the symmetric configuration, Charlie is equidistant from both parties. 
The maximum transmission distance for the ideal single-mode protocol is 7.04 km. 
When the mode-matching coefficients between Charlie and both parties are 95\% 
(i.e., $\eta_{\mathrm{m}}^{\mathrm{A}}=0.95$, $\eta_{\mathrm{m}}^{\mathrm{B}}=0.95$), 
the maximum transmission distance decreases to 4.8 km, 
which indicates a 31.8\% reduction in transmission distance compared to the ideal case.

\textbf{Case 2:} An extremely asymmetric configuration with Charlie located on Alice's 
side ($L_{AC}=0$). 
The impact of mode-matching coefficients on the secret key rate is illustrated in Fig.~\ref{C at mid, C at A}(b). 
In this configuration, the maximum transmission distance for the ideal single-mode protocol is 5.43 km. 
When the mode-matching 
coefficients decreases to 95\%, the maximum transmission distance is reduced to 3.49 km, 
which indicates a 35.7\% reduction in transmission distance compared to the ideal case.
\noindent
\begin{figure}[htbp]
    \includegraphics[width=0.48\textwidth]{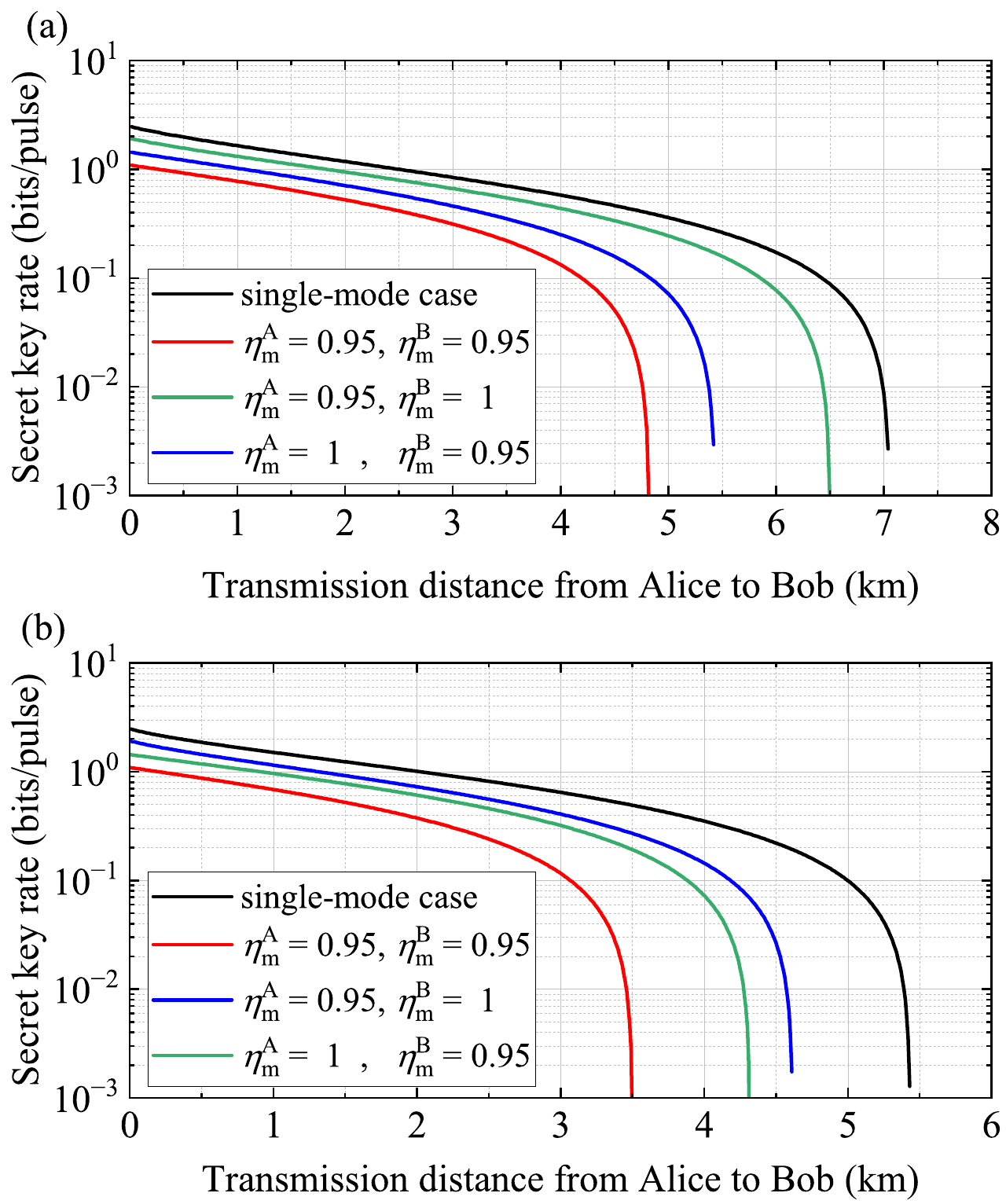} 
    \caption{
    \justifying 
    \label{C at mid, C at A} 
    (a) Symmetric structure of CV-MDI QKD. (b)Charlie placed at Alice's side.
    Black line represents the ideal single-mode case; green line represents $\eta_{\mathrm{m}}^{\mathrm{A}}=0.95$ and $\eta_{\mathrm{m}}^{\mathrm{B}}=1$; blue line represents $\eta_{\mathrm{m}}^{\mathrm{A}}=1$ and $\eta_{\mathrm{m}}^{\mathrm{B}}=0.95$; red line represents $\eta_{\mathrm{m}}^{\mathrm{A}}=\eta_{\mathrm{m}}^{\mathrm{B}}=0.95$.
    }
\end{figure}

\begin{figure}[h]
    \centering
    \includegraphics[width=0.48\textwidth]{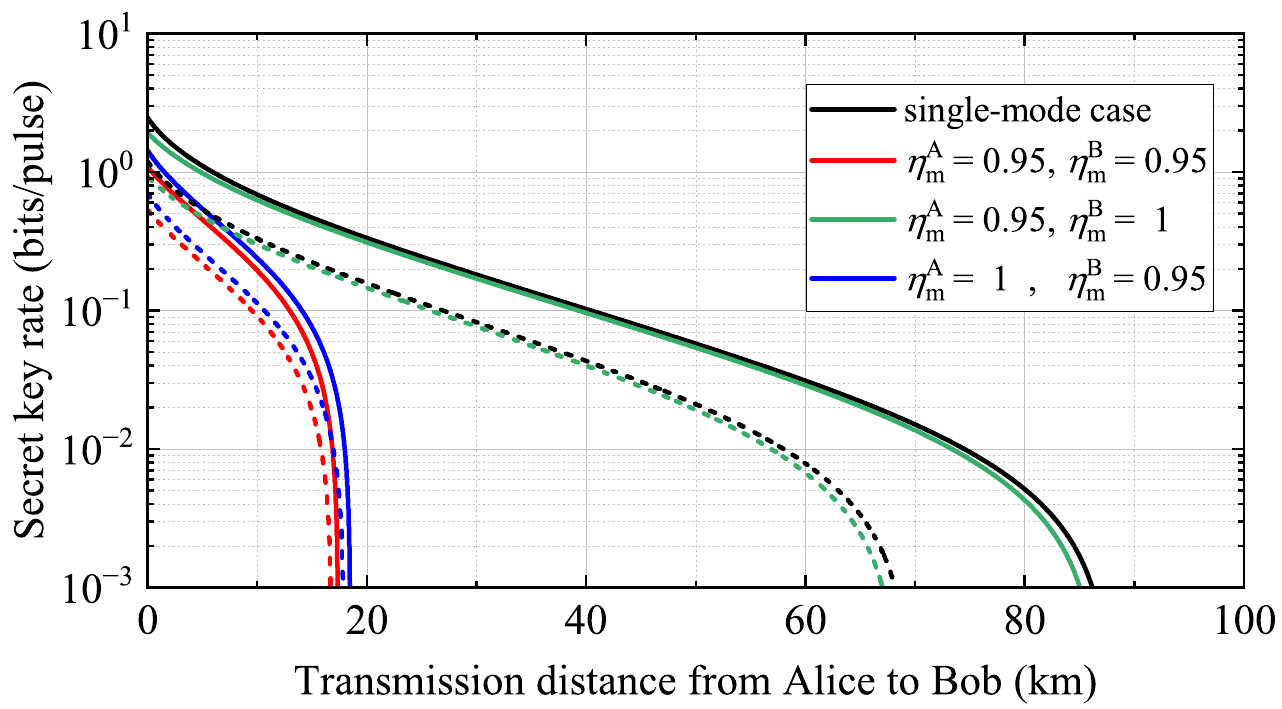}
    \caption{
    \justifying 
    Extremely asymmetric structure with Charlie placed at Bob's side.
    The dashed lines represent the results considering finite-size 
    effects with a code length of $N=10^{8}$. 
    Black line represents the ideal single-mode case; 
    green line represents $\eta_{\mathrm{m}}^{\mathrm{A}} =0.95$ and $\eta_{\mathrm{m}}^{\mathrm{B}}=1$; 
    blue line represents $\eta_{\mathrm{m}}^{\mathrm{A}}=1$ and $\eta_{\mathrm{m}}^{\mathrm{B}}=0.95$; 
    red line represents $\eta_{\mathrm{m}}^{\mathrm{A}}=\eta_{\mathrm{m}}^{\mathrm{B}}=0.95$.
    }
    \label{C at B}
\end{figure}

\textbf{Case 3:} An extremely asymmetric configuration with Charlie located on Bob's 
side ($L_{BC}=0$). 
The transmission distance of this protocol configuration far exceeds that 
of the symmetric configuration and the one with Charlie located on Alice's side. 
Previous studies suggest that this configuration can achieve optimal performance for long-distance CV-MDI QKD~\cite{26MDI_li2014continuous}. 
Figure.~\ref{C at B} presents the impact of different mode-matching coefficient settings on the secret key rate, represented by solid lines.

Compared to the performance of case 1 and case 2, the impact of different mode-matching coefficients on performance is more pronounced in case 3. 
Maximum transmission distances under different conditions are given in Table~\ref{table III. Maximum Transmission Distances Under Different Conditions}.
\begin{table}[h!]
    \centering
    \caption{Maximum Transmission Distances Under Different Conditions}
    \begin{tabular}{c|c}
    \hline
    \textbf{Condition} & \textbf{Max. Transmission Distance (km)} \\ \hline
    Ideal single-mode & 87.96 \\
    $\eta_{\mathrm{m}}^{\mathrm{A}} = 0.95$, $\eta_{\mathrm{m}}^{\mathrm{B}} = 1$ & 86.83 \\
    $\eta_{\mathrm{m}}^{\mathrm{A}} = 1$, $\eta_{\mathrm{m}}^{\mathrm{B}} = 0.95$ & 18.5 \\
    $\eta_{\mathrm{m}}^{\mathrm{A}} = \eta_{\mathrm{m}}^{\mathrm{B}} = 0.95$ & 17.38 \\
    \hline
    \end{tabular}
    \label{table III. Maximum Transmission Distances Under Different Conditions}
\end{table}
With ideal detection for mode $A_{\xi_A}$, a 5\% mismatch between mode $B_{\xi_B}$ and Charlie's TM reduces the maximum transmission distance by nearly 70 km. 
Conversely, with ideal detection for mode $B_{\xi_B}$, a 5\% mismatch between mode $A_{\xi_A}$ and Charlie's TM reduces the transmission distance by only 1.13 km.

The results indicate that the secret key rate of the continuous-mode CV-MDI QKD protocol is more significantly influenced by $\eta_{\mathrm{m}}^{\mathrm{B}}$. 
In other words, the matching degree between Bob's coherent state on $\xi_B$-TM and Charlie's $\Xi$-TM has a greater impact on the system's performance than that of Alice's. 

Furthermore, to better approximate practical conditions, we considered the impact 
of finite-size effects~\cite{45finitesize_leverrier2010finite,46finitesize_furrer2012continuous,zhang2017finite,lupo2018continuousYORK} 
on the performance of the continuous-mode CV-MDI QKD system. 
Finite-size effects influence the estimation of transmittance and excess noise 
due to the finite data length $N$. 
In our simulations, we set $N=10^8$. 
We investigated the impact of finite-size effects on CV-MDI QKD protocols in 
the continuous-mode scenario but do not involve a full composable finite-size proof. 
We focus on the influence of finite-size effects during the 
parameter estimation process. The parameter estimation primarily relies on the central limit 
theorem and the maximum likelihood estimation theorem, both of which have been extensively 
studied in existing works ~\cite{zhang2017finite, lupo2018continuousYORK}.
The simulation results are depicted by dashed lines in Fig.~\ref{C at B}. 
The dashed lines represent asymptotic curves with finite-size corrections.
For detailed calculations and results, please refer to Appendix~\ref{Appendix B. Finite-size}.

In the case of a mismatch between $\xi_B$-TM and $\Xi$-TM, the impact of finite-size effects on the maximum transmission distance is minimal, 
with a reduction of no more than 0.6 km. 
In the long-distance transmission scenario with ideal detection of mode $B_{\xi_B}$, finite-size effects reduce the maximum transmission distance by nearly 19 km, a decrease of about 21\% compared to the case with an infinite code length. 
Therefore, it is crucial to pay more attention to the impact of finite-size effects on 
performance in long-distance transmissions.

\subsection{\label{IV. B}Impact of mode-matching coefficients on secret key rate at fixed transmission distance}

\begin{figure}[htbp]
    \includegraphics[width=0.48\textwidth]{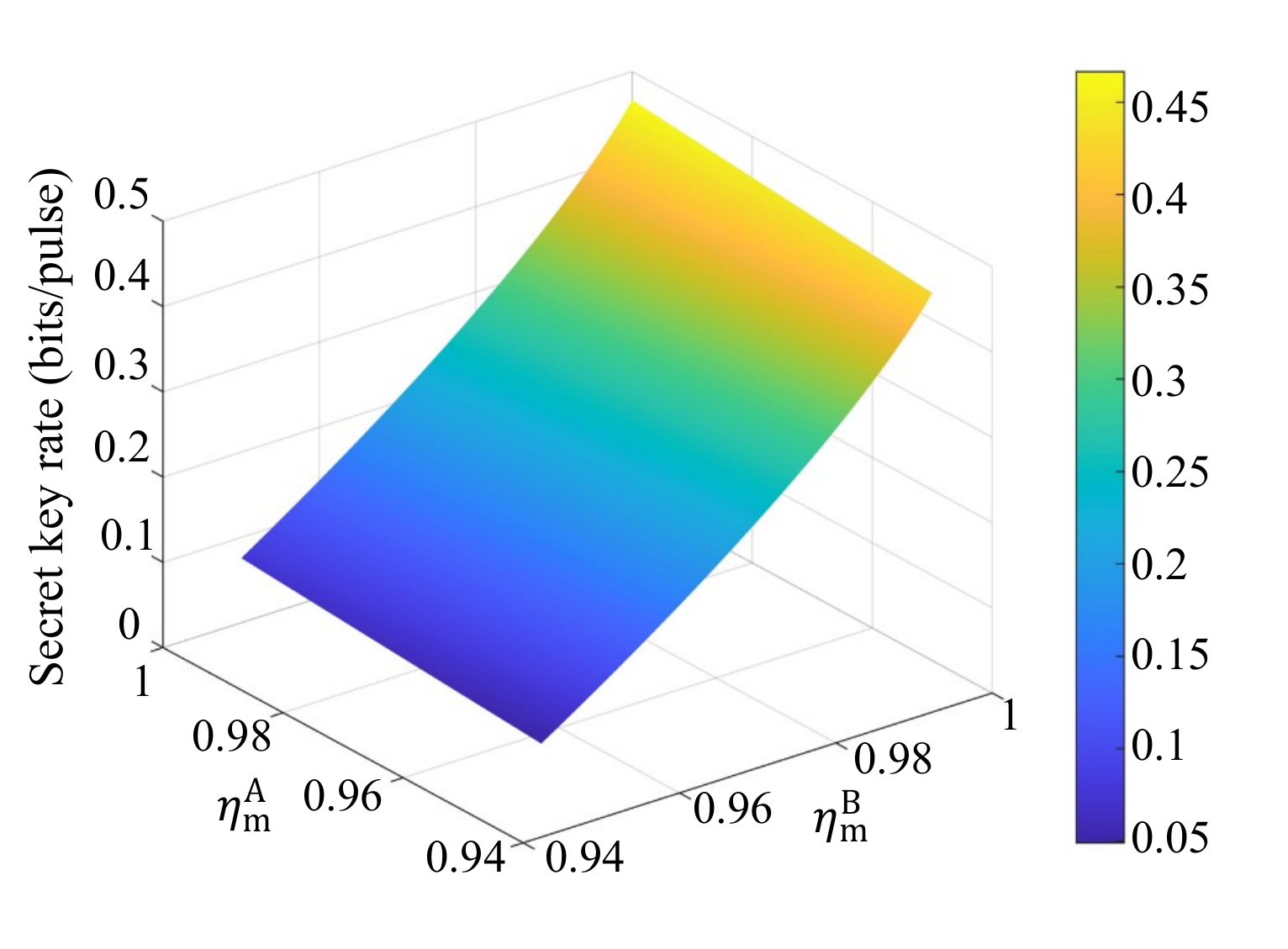} 
    \caption{
    \justifying 
    The secret key rate varies with changes in $\eta_{\mathrm{m}}^{\mathrm{A}}$ and $\eta_{\mathrm{m}}^{\mathrm{B}}$ at a fixed distance (15km).
    }
    \label{3D}
\end{figure}

To analyze the impact of mode-matching coefficients on the secret key rate at fixed transmission distances, we conducted a 3-dimensional simulation as depicted in Fig.~\ref{3D}. 
In this simulation, the secret key rate is significantly influenced by variations in $\eta_{\mathrm{m}}^{\mathrm{B}}$, while changes in $\eta_{\mathrm{m}}^{\mathrm{A}}$ have a less pronounced effect. 
For example, at a transmission distance of 15 km, when $\eta_{\mathrm{m}}^{\mathrm{A}}$ decreases by 5\%, the secret key rate decreases by 7.6\%. 
In contrast, when $\eta_{\mathrm{m}}^{\mathrm{B}}$ decreases by 5\%, the secret key rate drops by 83.8\%. 
Moreover, when $\eta_{\mathrm{m}}^{\mathrm{B}}$ decreases by more than 1\%, the secret key rate drops by over 20\%, significantly impacting system performance.

\noindent

Our explanation for this phenomenon is that, in the course of the protocol, Bob modifies his data based on Charlie's announced detection results, while Alice does not. 
Charlie's detection results are influenced 
by the mode matching between the received quantum states and his detectors.

Considering that Bob's data modification operation 
 is based on the measurement results announced by 
Charlie (Eve), its effect is essentially controlled by Charlie (Eve). 
Therefore, if we further assume that Bob's state preparation and displacement operations are 
entirely under Eve's control, leaving only the heterodyne detection on Bob's side as his own 
operation (as shown in Fig.~\ref{equivalent one-way}), the entire intermediate part controlled by Eve can be 
regarded as an insecure channel. This results in an equivalent one-way scenario ~\cite{26MDI_li2014continuous,8CVQKD_weedbrook2004quantum}, 
whose EB model is depicted in Fig.~\ref{equivalent one-way}.

More specifically, in the equivalent single-path model, 
the equivalent channel transmittance $T^{\prime}$ can be expressed as: $T^{\prime}=\frac{\eta_A}{2} \eta_{\mathrm{m}}^{\mathrm{A}} g^2$, 
the equivalent excess noise $\varepsilon^{\prime}$ can be expressed as: $\varepsilon^{\prime}=\varepsilon_A+\frac{1}{\eta_A}\left[\eta_B\left(\varepsilon_B-2 \eta_{\mathrm{m}}^{\mathrm{B}}\right)+2\right]$.
It can be observed that: the value of $\eta_{\mathrm{m}}^{\mathrm{B}}$ affects the trend of equivalent excess noise.

\begin{figure}[h]
    \includegraphics[width=0.48\textwidth]{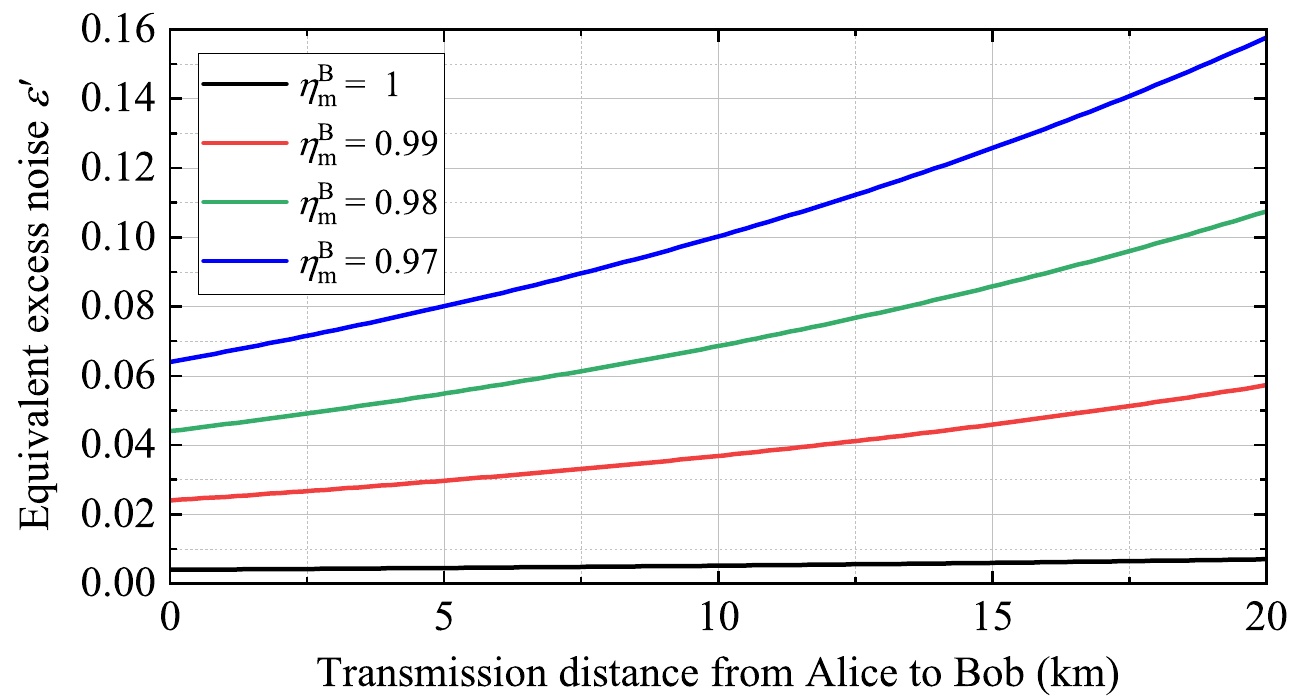} 
    \caption{
    \justifying 
    The impact of different $\eta_{\mathrm{m}}^{\mathrm{B}}$ on the equivalent excess noise.
    }
    \label{equivalent excess noise}
\end{figure}

Figure.~\ref{equivalent excess noise} shows the impact of different matching coefficients on the equivalent excess noise. 
As $\eta_{\mathrm{m}}^{\mathrm{B}}$ decreases, indicating a greater mismatch between Bob's coherent state on $\xi_B$-TM and Charlie's $\Xi$-TM, the equivalent excess noise at a fixed distance increases significantly. 
Additionally, as the mode-matching coefficient decreases, the equivalent excess noise becomes increasingly sensitive to changes in distance. 
As the transmission distance increases, the rise in equivalent excess noise becomes more pronounced, significantly exceeding the equivalent excess noise in the ideal single-mode case, thereby reducing system performance. 
This insight guides the design of CV-MDI QKD systems, emphasizing the importance of mode matching between the quantum states' TMs and the detectors' TMs. 
Specifically, maximizing the mode-matching coefficients between Bob's quantum state and the detectors can reduce equivalent excess noise, thereby optimizing system performance. 

\noindent

\section{\label{V. CONCLUSION}CONCLUSION}
In this paper, we analyzed the performance of CV-MDI QKD under continuous-mode scenarios by introducing TMs. 
The numerical simulations indicated that the mismatch between the senders' TMs and the Bell measurement TM leads to a reduction in the maximum transmission distance and secret key rate.
Due to the asymmetry in the data modification step, the system's performance is 
more sensitive to the mismatch between Bob's transmitting mode and the Bell measurement mode. 
When Charlie is close to Bob, even a 5\% mismatch reduces the transmission distance 
from 87.96 km to 18.50 km, and at 15 km, the key rate drops by more than 80\% compared to the ideal case.
This mismatch requires careful attention due to its significant impact on system performance. 
Precise calibration of Bob's mode is needed to mitigate this impact.

By introducing TMs as an analytical tool, we can address previously challenging scenarios, 
such as the interference between two broadband optical sources studied in this work, 
while also extending to broader application scenarios. For instance, this approach provides 
guidance for the selection and optimization of device parameters in experiments under 
different conditions. Furthermore, it offers critical theoretical insights for calibrating 
and optimizing multi-user interference in future large-scale MDI networks.
In such networks, the mode-matching coefficients between each pair of nodes 
should be carefully considered to optimize overall network performance. 
Our method provides a potential means to analyze these mode-matching effects.

\section*{Acknowledgments}
This work was supported by the National Natural Science Foundation of China under Grant No. 62371060, No. 62001041, No. 62201012, the Fund of State Key Laboratory of Information Photonics and Optical Communications under Grant No. IPOC2022ZT09. 

%

\appendix
\section{\label{Appendix A. quadratures}Relationship of quadratures used in numerical simulation}
Considering the mode-matching coefficients between the quantum states of the senders and detector 1, the modes, after passing through the quantum channel, become:
\begin{equation}
    \text{\small
        $\begin{array}{l}
        \hat{A}_{\Xi}^{\prime} = \sqrt{\eta_A}\left(\sqrt{\eta_{\mathrm{m}}^{\mathrm{A} 1}} \hat{A}_2^{\xi_A} + \sqrt{1-\eta_{\mathrm{m}}^{\mathrm{A} 1}} \hat{A}_2^{\Psi_{\perp}}\right) + \sqrt{1-\eta_A} \hat{E}_2, \\
        \hat{B}_{\Xi}^{\prime} = \sqrt{\eta_B}\left(\sqrt{\eta_{\mathrm{m}}^{\mathrm{B} 1}} \hat{B}_2^{\xi_B} + \sqrt{1-\eta_{\mathrm{m}}^{\mathrm{B} 1}} \hat{B}_2^{\Psi_{\perp}}\right) + \sqrt{1-\eta_B} \hat{E}_5.
        \end{array}$
    }
\end{equation}
    
Considering the mode-matching coefficients between the quantum states of the senders and detector 2, the modes, after passing through the quantum channel, become:
\begin{equation}
    \text{\small
        $\begin{array}{l}
        \hat{A}_{\Xi}^{\prime} = \sqrt{\eta_A}\left(\sqrt{\eta_{\mathrm{m}}^{\mathrm{A} 2}} \hat{A}_2^{\xi_A} + \sqrt{1-\eta_{\mathrm{m}}^{\mathrm{A} 2}} \hat{A}_2^{\Psi_{\perp}}\right) + \sqrt{1-\eta_A} \hat{E}_2, \\
        \hat{B}_{\Xi}^{\prime} = \sqrt{\eta_B}\left(\sqrt{\eta_{\mathrm{m}}^{\mathrm{B} 2}} \hat{B}_2^{\xi_B} + \sqrt{1-\eta_{\mathrm{m}}^{\mathrm{B} 2}} \hat{B}_2^{\Psi_{\perp}}\right) + \sqrt{1-\eta_B} \hat{E}_5.
        \end{array}$
    }
\end{equation}

Modes ${A}_{\Xi}^{\prime}$ and ${B}_{\Xi}^{\prime}$ interfere on the 50:50 BS, 
then modes ${C}_{\Xi}$ and ${D}_{\Xi}$ are
\begin{equation}
    \text{\small
        $\begin{aligned}
            \hat{C}_{\Xi} &= \frac{1}{\sqrt{2}}\left(\hat{A}_{\Xi}^{\prime} - \hat{B}_{\Xi}^{\prime}\right) \\
            &= \frac{1}{\sqrt{2}}\left(\sqrt{\eta_A \eta_{\mathrm{m}}^{\mathrm{A} 1}} \hat{A}_2^{\xi_A} - \sqrt{\eta_B \eta_{\mathrm{m}}^{\mathrm{B} 1}} \hat{B}_2^{\xi_B}\right) \\
            &+ \frac{1}{\sqrt{2}}\left(\sqrt{\eta_A (1 - \eta_{\mathrm{m}}^{\mathrm{A} 1})} \hat{A}_2^{\Psi_{\perp}} - \sqrt{\eta_B (1 - \eta_{\mathrm{m}}^{\mathrm{B} 1})} \hat{B}_2^{\Psi_{\perp}}\right) \\
            &+ \frac{1}{\sqrt{2}}\left(\sqrt{1 - \eta_A} \hat{E}_2 - \sqrt{1 - \eta_B} \hat{E}_5\right)
        \end{aligned}$
    }
\end{equation}
and
\begin{equation}
    \text{\small
        $\begin{aligned}
            \hat{D}_{\Xi} &= \frac{1}{\sqrt{2}}\left(\hat{A}_{\Xi}^{\prime} + \hat{B}_{\Xi}^{\prime}\right) \\
            &= \frac{1}{\sqrt{2}}\left(\sqrt{\eta_A \eta_{\mathrm{m}}^{\mathrm{A} 2}} \hat{A}_2^{\xi_A} + \sqrt{\eta_B \eta_{\mathrm{m}}^{\mathrm{B} 2}} \hat{B}_2^{\xi_B}\right) \\
            &+ \frac{1}{\sqrt{2}}\left(\sqrt{\eta_A (1 - \eta_{\mathrm{m}}^{\mathrm{A} 2})} \hat{A}_2^{\Psi_{\perp}} + \sqrt{\eta_B (1 - \eta_{\mathrm{m}}^{\mathrm{B} 2})} \hat{B}_2^{\Psi_{\perp}}\right) \\
            &+ \frac{1}{\sqrt{2}}\left(\sqrt{1 - \eta_A} \hat{E}_2 + \sqrt{1 - \eta_B} \hat{E}_5\right).
        \end{aligned}$
    }
\end{equation}

After the displacement, mode $B_1^{\prime}$ becomes
\begin{equation}
    \text{\small
        $\begin{aligned}
            \hat{B}^{\prime}{ }_{1 x}^{\xi_B} &= \hat{B}_{1 x}^{\xi_B} + g_m \hat{C}_x^{\Xi} \\
            &= \left(\hat{B}_{1 x}^{\xi_B} - \frac{g_m}{\sqrt{2}} \sqrt{\eta_B \eta_{\mathrm{m}}^{\mathrm{B} 1}} \hat{B}_{2 x}^{\xi_B}\right) \\
            &\quad + \frac{g_m}{\sqrt{2}} \sqrt{\eta_A \eta_{\mathrm{m}}^{\mathrm{A} 1}} \hat{A}_{2 x}^{\xi_A} \\
            &\quad + \frac{g_m}{\sqrt{2}}\left(\sqrt{1 - \eta_A} \hat{E}_{2 x} - \sqrt{1 - \eta_B} \hat{E}_{5 x}\right) \\
            &\quad + \frac{g_m}{\sqrt{2}}\left(\sqrt{\eta_A (1 - \eta_{\mathrm{m}}^{\mathrm{A} 1})} \hat{A}_{2x}^{\Psi_{\perp}} - \sqrt{\eta_B (1 - \eta_{\mathrm{m}}^{\mathrm{B} 1})} \hat{B}_{2x}^{\Psi_{\perp}}\right)
        \end{aligned}$
    }
\end{equation}
and
\begin{equation}
    \text{\small
        $\begin{aligned}
            \hat{B}^{\prime}{ }_{1 p}^{\xi_B} &= \hat{B}_{1 p}^{\xi_B} + g_m \hat{D}_p^{\Xi} \\
            &= \left(\hat{B}_{1 p}^{\xi_B} + \frac{g_m}{\sqrt{2}} \sqrt{\eta_B \eta_{\mathrm{m}}^{\mathrm{B} 2}} \hat{B}_{2 p}^{\xi_B}\right) \\
            &\quad + \frac{g_m}{\sqrt{2}} \sqrt{\eta_A \eta_{\mathrm{m}}^{\mathrm{A} 2}} \hat{A}_{2 p}^{\xi_A} \\
            &\quad + \frac{g_m}{\sqrt{2}}\left(\sqrt{1 - \eta_A} \hat{E}_{2 p} + \sqrt{1 - \eta_B} \hat{E}_{5 p}\right) \\
            &\quad + \frac{g_m}{\sqrt{2}}\left(\sqrt{\eta_A (1 - \eta_{\mathrm{m}}^{\mathrm{A} 2})} \hat{A}_{2p}^{\Psi_{\perp}} + \sqrt{\eta_B (1 - \eta_{\mathrm{m}}^{\mathrm{B} 2})} \hat{B}_{2p}^{\Psi_{\perp}}\right).
        \end{aligned}$
    }
\end{equation}

\section{\label{Appendix B. Finite-size}Finite-size analysis of CV-MDI QKD under continuous-mode scenario}
Considering finite-size effects, Alice and Bob need to select $m$ signals 
from the exchanged $N$ signals for parameter estimation, leaving $n=N-m$ 
signals for key generation. 
The secret key rate formula is modified to~\cite{45finitesize_leverrier2010finite}:
\begin{equation}
    K_{\text {finite }}=\frac{n}{N}[\beta_R I(A: B)-S(E: B)-\Delta(n)],
\end{equation}
where $\Delta(n)$ is associated with the security of privacy amplification, and its value is given by:
\begin{equation}
    \Delta(n)=\left(2 \operatorname{dim} H_X+3\right) \sqrt{\frac{\log _2(2 / \bar{\epsilon})}{n}}+\frac{2}{n} \log _2 \frac{1}{\epsilon_{P A}}.
\end{equation}
$H_X$ represents the Hilbert space dimension of the variable $x$ in the raw key. 
In the CV protocol, the $\operatorname{dim} H_X$ is set to 2. 
The smoothing parameter $\bar{\epsilon}$ and the privacy amplification parameter $\epsilon_{P A}$ are intermediate variables, with their optimal values set to $\bar{\epsilon} = \epsilon_{P A} = 10^{-10}$.

The Alice-Charlie channel and Bob-Charlie channel are Gaussian channels, with the data relationships given by:
\begin{equation}
    \begin{array}{l}
        y_A^{\prime} = t_A^{\prime} x_A + z_A, \\
        y_B^{\prime} = t_B^{\prime} x_B + z_B,
    \end{array}
\end{equation}
where $t_A^{\prime}=\sqrt{\eta_A}$, $t_B^{\prime}=\sqrt{\eta_B}$. 
$Z_A$ and $Z_B$ are normally distributed with variances 
$\sigma_A^{\prime 2}=1+\eta_A \varepsilon_A$ and 
$\sigma_B^{\prime 2}=1+\eta_B \varepsilon_B$, respectively.
To ensure the security of the protocol, the channel transmittance $\eta_A$ and $\eta_B$ should be minimized, while the channel excess noise $\varepsilon_A$ and $\varepsilon_B$ should be maximized.
By substituting the parameters to be estimated into Eq.~(\ref{covariance matrix}) from Section III, matrix $\gamma_{f}$ is obtained:
\begin{equation}
    \gamma_{f}=\left[\begin{array}{cc}V_A I & \sqrt{t_{\text {min }} \eta_{\mathrm{m}}^{\mathrm{A}}\left(V_A^2-1\right)} \sigma_z \\ \sqrt{t_{\text {min }} \eta_{\mathrm{m}}^{\mathrm{A}}\left(V_A^2-1\right)} \sigma_z & V_{B_1^{\prime}} I\end{array}\right],
\end{equation}
where $V_{B_1^{\prime}}$=$(t_{\min }){ }^2 \eta_{\mathrm{m}}^{\mathrm{A}}\left(V_A-1\right)+1+t_{\min } \varepsilon_{\max }^{\prime}$.

\begin{equation}
    t_{\min }=\frac{t_{A_{\min }}^{\prime}}{t_{B_{\min }}^{\prime}} \frac{V_B-1}{V_B+1} \frac{1}{\eta_{\mathrm{m}}^{\mathrm{B}}}.
\end{equation}

\begin{equation}
    t_{\min } \varepsilon_{\max }^{\prime}=\frac{1}{\eta_{\mathrm{m}}^{\mathrm{B}}} \frac{V_B-1}{V_B+1} 
    \frac{(\sigma_{A_{\max }}^{\prime})^2+(\sigma_{B_{\max }}^{\prime})^2-2 (t_{B_{\max }}^{\prime})^2}{(t_{B_{\max }}^{\prime})^2}.
\end{equation}




%


\begin{thebibliography}{69}%
\makeatletter
\providecommand \@ifxundefined [1]{%
 \@ifx{#1\undefined}
}%
\providecommand \@ifnum [1]{%
 \ifnum #1\expandafter \@firstoftwo
 \else \expandafter \@secondoftwo
 \fi
}%
\providecommand \@ifx [1]{%
 \ifx #1\expandafter \@firstoftwo
 \else \expandafter \@secondoftwo
 \fi
}%
\providecommand \natexlab [1]{#1}%
\providecommand \enquote  [1]{``#1''}%
\providecommand \bibnamefont  [1]{#1}%
\providecommand \bibfnamefont [1]{#1}%
\providecommand \citenamefont [1]{#1}%
\providecommand \href@noop [0]{\@secondoftwo}%
\providecommand \href [0]{\begingroup \@sanitize@url \@href}%
\providecommand \@href[1]{\@@startlink{#1}\@@href}%
\providecommand \@@href[1]{\endgroup#1\@@endlink}%
\providecommand \@sanitize@url [0]{\catcode `\\12\catcode `\$12\catcode
  `\&12\catcode `\#12\catcode `\^12\catcode `\_12\catcode `\%12\relax}%
\providecommand \@@startlink[1]{}%
\providecommand \@@endlink[0]{}%
\providecommand \url  [0]{\begingroup\@sanitize@url \@url }%
\providecommand \@url [1]{\endgroup\@href {#1}{\urlprefix }}%
\providecommand \urlprefix  [0]{URL }%
\providecommand \Eprint [0]{\href }%
\providecommand \doibase [0]{https://doi.org/}%
\providecommand \selectlanguage [0]{\@gobble}%
\providecommand \bibinfo  [0]{\@secondoftwo}%
\providecommand \bibfield  [0]{\@secondoftwo}%
\providecommand \translation [1]{[#1]}%
\providecommand \BibitemOpen [0]{}%
\providecommand \bibitemStop [0]{}%
\providecommand \bibitemNoStop [0]{.\EOS\space}%
\providecommand \EOS [0]{\spacefactor3000\relax}%
\providecommand \BibitemShut  [1]{\csname bibitem#1\endcsname}%
\let\auto@bib@innerbib\@empty
\bibitem [{\citenamefont {Bennett}\ and\ \citenamefont
  {Brassard}(2014)}]{bennett2014quantum}%
  \BibitemOpen
  \bibfield  {author} {\bibinfo {author} {\bibfnamefont {C.~H.}\ \bibnamefont
  {Bennett}}\ and\ \bibinfo {author} {\bibfnamefont {G.}~\bibnamefont
  {Brassard}},\ }\bibinfo {title} {Quantum cryptography: Public key
  distribution and coin tossing},\ \href@noop {} {\bibfield  {journal}
  {\bibinfo  {journal} {Theoretical computer science}\ }\textbf {\bibinfo
  {volume} {560}},\ \bibinfo {pages} {7} (\bibinfo {year} {2014})}\BibitemShut
  {NoStop}%
\bibitem [{\citenamefont {Ekert}(1991)}]{1ekert1991quantum}%
  \BibitemOpen
  \bibfield  {author} {\bibinfo {author} {\bibfnamefont {A.~K.}\ \bibnamefont
  {Ekert}},\ }\bibinfo {title} {Quantum cryptography based on bell's theorem},\
  \href@noop {} {\bibfield  {journal} {\bibinfo  {journal} {Phys. Rev. Lett.}\
  }\textbf {\bibinfo {volume} {67}},\ \bibinfo {pages} {661} (\bibinfo {year}
  {1991})}\BibitemShut {NoStop}%
\bibitem [{\citenamefont {Gisin}\ \emph {et~al.}(2002)\citenamefont {Gisin},
  \citenamefont {Ribordy}, \citenamefont {Tittel},\ and\ \citenamefont
  {Zbinden}}]{2QKD_gisin2002quantum}%
  \BibitemOpen
  \bibfield  {author} {\bibinfo {author} {\bibfnamefont {N.}~\bibnamefont
  {Gisin}}, \bibinfo {author} {\bibfnamefont {G.}~\bibnamefont {Ribordy}},
  \bibinfo {author} {\bibfnamefont {W.}~\bibnamefont {Tittel}},\ and\ \bibinfo
  {author} {\bibfnamefont {H.}~\bibnamefont {Zbinden}},\ }\bibinfo {title}
  {Quantum cryptography},\ \href@noop {} {\bibfield  {journal} {\bibinfo
  {journal} {Rev. Mod. Phys.}\ }\textbf {\bibinfo {volume} {74}},\ \bibinfo
  {pages} {145} (\bibinfo {year} {2002})}\BibitemShut {NoStop}%
\bibitem [{\citenamefont {Pirandola}\ \emph {et~al.}(2020)\citenamefont
  {Pirandola}, \citenamefont {Andersen}, \citenamefont {Banchi}, \citenamefont
  {Berta}, \citenamefont {Bunandar}, \citenamefont {Colbeck}, \citenamefont
  {Englund}, \citenamefont {Gehring}, \citenamefont {Lupo}, \citenamefont
  {Ottaviani} \emph {et~al.}}]{3QKD_pirandola2020advances}%
  \BibitemOpen
  \bibfield  {author} {\bibinfo {author} {\bibfnamefont {S.}~\bibnamefont
  {Pirandola}}, \bibinfo {author} {\bibfnamefont {U.~L.}\ \bibnamefont
  {Andersen}}, \bibinfo {author} {\bibfnamefont {L.}~\bibnamefont {Banchi}},
  \bibinfo {author} {\bibfnamefont {M.}~\bibnamefont {Berta}}, \bibinfo
  {author} {\bibfnamefont {D.}~\bibnamefont {Bunandar}}, \bibinfo {author}
  {\bibfnamefont {R.}~\bibnamefont {Colbeck}}, \bibinfo {author} {\bibfnamefont
  {D.}~\bibnamefont {Englund}}, \bibinfo {author} {\bibfnamefont
  {T.}~\bibnamefont {Gehring}}, \bibinfo {author} {\bibfnamefont
  {C.}~\bibnamefont {Lupo}}, \bibinfo {author} {\bibfnamefont {C.}~\bibnamefont
  {Ottaviani}}, \emph {et~al.},\ }\bibinfo {title} {Advances in quantum
  cryptography},\ \href@noop {} {\bibfield  {journal} {\bibinfo  {journal}
  {Adv. Opt. Photonics}\ }\textbf {\bibinfo {volume} {12}},\ \bibinfo {pages}
  {1012} (\bibinfo {year} {2020})}\BibitemShut {NoStop}%
\bibitem [{\citenamefont {Xu}\ \emph {et~al.}(2020)\citenamefont {Xu},
  \citenamefont {Ma}, \citenamefont {Zhang}, \citenamefont {Lo},\ and\
  \citenamefont {Pan}}]{4QKD_xu2020secure}%
  \BibitemOpen
  \bibfield  {author} {\bibinfo {author} {\bibfnamefont {F.}~\bibnamefont
  {Xu}}, \bibinfo {author} {\bibfnamefont {X.}~\bibnamefont {Ma}}, \bibinfo
  {author} {\bibfnamefont {Q.}~\bibnamefont {Zhang}}, \bibinfo {author}
  {\bibfnamefont {H.-K.}\ \bibnamefont {Lo}},\ and\ \bibinfo {author}
  {\bibfnamefont {J.-W.}\ \bibnamefont {Pan}},\ }\bibinfo {title} {Secure
  quantum key distribution with realistic devices},\ \href@noop {} {\bibfield
  {journal} {\bibinfo  {journal} {Rev. Mod. Phys.}\ }\textbf {\bibinfo {volume}
  {92}},\ \bibinfo {pages} {025002} (\bibinfo {year} {2020})}\BibitemShut
  {NoStop}%
\bibitem [{\citenamefont {Shannon}(1949)}]{5OTP_shannon1949communication}%
  \BibitemOpen
  \bibfield  {author} {\bibinfo {author} {\bibfnamefont {C.~E.}\ \bibnamefont
  {Shannon}},\ }\bibinfo {title} {Communication theory of secrecy systems},\
  \href@noop {} {\bibfield  {journal} {\bibinfo  {journal} {Bell Syst. Tech.
  J.}\ }\textbf {\bibinfo {volume} {28}},\ \bibinfo {pages} {656} (\bibinfo
  {year} {1949})}\BibitemShut {NoStop}%
\bibitem [{\citenamefont {Shor}(1994)}]{6shor1994algorithms}%
  \BibitemOpen
  \bibfield  {author} {\bibinfo {author} {\bibfnamefont {P.~W.}\ \bibnamefont
  {Shor}},\ }\bibinfo {title} {Algorithms for quantum computation: discrete
  logarithms and factoring},\ in\ \href@noop {} {\emph {\bibinfo {booktitle}
  {Proceedings 35th annual symposium on foundations of computer science}}}\
  (\bibinfo {organization} {Ieee},\ \bibinfo {year} {1994})\ pp.\ \bibinfo
  {pages} {124--134}\BibitemShut {NoStop}%
\bibitem [{\citenamefont {Grosshans}\ and\ \citenamefont
  {Grangier}(2002)}]{7CVQKD_grosshans2002continuous}%
  \BibitemOpen
  \bibfield  {author} {\bibinfo {author} {\bibfnamefont {F.}~\bibnamefont
  {Grosshans}}\ and\ \bibinfo {author} {\bibfnamefont {P.}~\bibnamefont
  {Grangier}},\ }\bibinfo {title} {Continuous variable quantum cryptography
  using coherent states},\ \href@noop {} {\bibfield  {journal} {\bibinfo
  {journal} {Phys. Rev. Lett.}\ }\textbf {\bibinfo {volume} {88}},\ \bibinfo
  {pages} {057902} (\bibinfo {year} {2002})}\BibitemShut {NoStop}%
\bibitem [{\citenamefont {Weedbrook}\ \emph {et~al.}(2004)\citenamefont
  {Weedbrook}, \citenamefont {Lance}, \citenamefont {Bowen}, \citenamefont
  {Symul}, \citenamefont {Ralph},\ and\ \citenamefont
  {Lam}}]{8CVQKD_weedbrook2004quantum}%
  \BibitemOpen
  \bibfield  {author} {\bibinfo {author} {\bibfnamefont {C.}~\bibnamefont
  {Weedbrook}}, \bibinfo {author} {\bibfnamefont {A.~M.}\ \bibnamefont
  {Lance}}, \bibinfo {author} {\bibfnamefont {W.~P.}\ \bibnamefont {Bowen}},
  \bibinfo {author} {\bibfnamefont {T.}~\bibnamefont {Symul}}, \bibinfo
  {author} {\bibfnamefont {T.~C.}\ \bibnamefont {Ralph}},\ and\ \bibinfo
  {author} {\bibfnamefont {P.~K.}\ \bibnamefont {Lam}},\ }\bibinfo {title}
  {Quantum cryptography without switching},\ \href@noop {} {\bibfield
  {journal} {\bibinfo  {journal} {Phys. Rev. Lett.}\ }\textbf {\bibinfo
  {volume} {93}},\ \bibinfo {pages} {170504} (\bibinfo {year}
  {2004})}\BibitemShut {NoStop}%
\bibitem [{\citenamefont {Leverrier}\ \emph {et~al.}(2013)\citenamefont
  {Leverrier}, \citenamefont {Garc{\'\i}a-Patr{\'o}n}, \citenamefont {Renner},\
  and\ \citenamefont {Cerf}}]{9leverrier2013security}%
  \BibitemOpen
  \bibfield  {author} {\bibinfo {author} {\bibfnamefont {A.}~\bibnamefont
  {Leverrier}}, \bibinfo {author} {\bibfnamefont {R.}~\bibnamefont
  {Garc{\'\i}a-Patr{\'o}n}}, \bibinfo {author} {\bibfnamefont {R.}~\bibnamefont
  {Renner}},\ and\ \bibinfo {author} {\bibfnamefont {N.~J.}\ \bibnamefont
  {Cerf}},\ }\bibinfo {title} {Security of continuous-variable quantum key
  distribution against general attacks},\ \href@noop {} {\bibfield  {journal}
  {\bibinfo  {journal} {Phys. Rev. Lett.}\ }\textbf {\bibinfo {volume} {110}},\
  \bibinfo {pages} {030502} (\bibinfo {year} {2013})}\BibitemShut {NoStop}%
\bibitem [{\citenamefont {Leverrier}(2015)}]{10leverrier2015composable}%
  \BibitemOpen
  \bibfield  {author} {\bibinfo {author} {\bibfnamefont {A.}~\bibnamefont
  {Leverrier}},\ }\bibinfo {title} {Composable security proof for
  continuous-variable quantum key distribution with coherent states},\
  \href@noop {} {\bibfield  {journal} {\bibinfo  {journal} {Phys. Rev. Lett.}\
  }\textbf {\bibinfo {volume} {114}},\ \bibinfo {pages} {070501} (\bibinfo
  {year} {2015})}\BibitemShut {NoStop}%
\bibitem [{\citenamefont {Leverrier}(2017)}]{11leverrier2017security}%
  \BibitemOpen
  \bibfield  {author} {\bibinfo {author} {\bibfnamefont {A.}~\bibnamefont
  {Leverrier}},\ }\bibinfo {title} {Security of continuous-variable quantum key
  distribution via a gaussian de finetti reduction},\ \href@noop {} {\bibfield
  {journal} {\bibinfo  {journal} {Phys. Rev. Lett.}\ }\textbf {\bibinfo
  {volume} {118}},\ \bibinfo {pages} {200501} (\bibinfo {year}
  {2017})}\BibitemShut {NoStop}%
\bibitem [{\citenamefont {Pirandola}\ \emph {et~al.}(2017)\citenamefont
  {Pirandola}, \citenamefont {Laurenza}, \citenamefont {Ottaviani},\ and\
  \citenamefont {Banchi}}]{12pirandola2017fundamental}%
  \BibitemOpen
  \bibfield  {author} {\bibinfo {author} {\bibfnamefont {S.}~\bibnamefont
  {Pirandola}}, \bibinfo {author} {\bibfnamefont {R.}~\bibnamefont {Laurenza}},
  \bibinfo {author} {\bibfnamefont {C.}~\bibnamefont {Ottaviani}},\ and\
  \bibinfo {author} {\bibfnamefont {L.}~\bibnamefont {Banchi}},\ }\bibinfo
  {title} {Fundamental limits of repeaterless quantum communications},\
  \href@noop {} {\bibfield  {journal} {\bibinfo  {journal} {Nat. Commun.}\
  }\textbf {\bibinfo {volume} {8}},\ \bibinfo {pages} {1} (\bibinfo {year}
  {2017})}\BibitemShut {NoStop}%
\bibitem [{\citenamefont {Ghorai}\ \emph {et~al.}(2019)\citenamefont {Ghorai},
  \citenamefont {Grangier}, \citenamefont {Diamanti},\ and\ \citenamefont
  {Leverrier}}]{ghorai2019asymptotic}%
  \BibitemOpen
  \bibfield  {author} {\bibinfo {author} {\bibfnamefont {S.}~\bibnamefont
  {Ghorai}}, \bibinfo {author} {\bibfnamefont {P.}~\bibnamefont {Grangier}},
  \bibinfo {author} {\bibfnamefont {E.}~\bibnamefont {Diamanti}},\ and\
  \bibinfo {author} {\bibfnamefont {A.}~\bibnamefont {Leverrier}},\ }\bibinfo
  {title} {Asymptotic security of continuous-variable quantum key distribution
  with a discrete modulation},\ \href@noop {} {\bibfield  {journal} {\bibinfo
  {journal} {Phys. Rev. X}\ }\textbf {\bibinfo {volume} {9}},\ \bibinfo {pages}
  {021059} (\bibinfo {year} {2019})}\BibitemShut {NoStop}%
\bibitem [{\citenamefont {Lin}\ \emph {et~al.}(2019)\citenamefont {Lin},
  \citenamefont {Upadhyaya},\ and\ \citenamefont
  {L{\"u}tkenhaus}}]{lin2019asymptotic}%
  \BibitemOpen
  \bibfield  {author} {\bibinfo {author} {\bibfnamefont {J.}~\bibnamefont
  {Lin}}, \bibinfo {author} {\bibfnamefont {T.}~\bibnamefont {Upadhyaya}},\
  and\ \bibinfo {author} {\bibfnamefont {N.}~\bibnamefont {L{\"u}tkenhaus}},\
  }\bibinfo {title} {Asymptotic security analysis of discrete-modulated
  continuous-variable quantum key distribution},\ \href@noop {} {\bibfield
  {journal} {\bibinfo  {journal} {Phys. Rev. X}\ }\textbf {\bibinfo {volume}
  {9}},\ \bibinfo {pages} {041064} (\bibinfo {year} {2019})}\BibitemShut
  {NoStop}%
\bibitem [{\citenamefont {Upadhyaya}\ \emph {et~al.}(2021)\citenamefont
  {Upadhyaya}, \citenamefont {van Himbeeck}, \citenamefont {Lin},\ and\
  \citenamefont {L{\"u}tkenhaus}}]{upadhyaya2021dimension}%
  \BibitemOpen
  \bibfield  {author} {\bibinfo {author} {\bibfnamefont {T.}~\bibnamefont
  {Upadhyaya}}, \bibinfo {author} {\bibfnamefont {T.}~\bibnamefont {van
  Himbeeck}}, \bibinfo {author} {\bibfnamefont {J.}~\bibnamefont {Lin}},\ and\
  \bibinfo {author} {\bibfnamefont {N.}~\bibnamefont {L{\"u}tkenhaus}},\
  }\bibinfo {title} {Dimension reduction in quantum key distribution for
  continuous-and discrete-variable protocols},\ \href@noop {} {\bibfield
  {journal} {\bibinfo  {journal} {PRX Quantum}\ }\textbf {\bibinfo {volume}
  {2}},\ \bibinfo {pages} {020325} (\bibinfo {year} {2021})}\BibitemShut
  {NoStop}%
\bibitem [{\citenamefont {Denys}\ \emph {et~al.}(2021)\citenamefont {Denys},
  \citenamefont {Brown},\ and\ \citenamefont {Leverrier}}]{denys2021explicit}%
  \BibitemOpen
  \bibfield  {author} {\bibinfo {author} {\bibfnamefont {A.}~\bibnamefont
  {Denys}}, \bibinfo {author} {\bibfnamefont {P.}~\bibnamefont {Brown}},\ and\
  \bibinfo {author} {\bibfnamefont {A.}~\bibnamefont {Leverrier}},\ }\bibinfo
  {title} {Explicit asymptotic secret key rate of continuous-variable quantum
  key distribution with an arbitrary modulation},\ \href@noop {} {\bibfield
  {journal} {\bibinfo  {journal} {Quantum}\ }\textbf {\bibinfo {volume} {5}},\
  \bibinfo {pages} {540} (\bibinfo {year} {2021})}\BibitemShut {NoStop}%
\bibitem [{\citenamefont {Lupo}\ and\ \citenamefont
  {Ouyang}(2022)}]{lupo2022quantum}%
  \BibitemOpen
  \bibfield  {author} {\bibinfo {author} {\bibfnamefont {C.}~\bibnamefont
  {Lupo}}\ and\ \bibinfo {author} {\bibfnamefont {Y.}~\bibnamefont {Ouyang}},\
  }\bibinfo {title} {Quantum key distribution with nonideal heterodyne
  detection: composable security of discrete-modulation continuous-variable
  protocols},\ \href@noop {} {\bibfield  {journal} {\bibinfo  {journal} {PRX
  Quantum}\ }\textbf {\bibinfo {volume} {3}},\ \bibinfo {pages} {010341}
  (\bibinfo {year} {2022})}\BibitemShut {NoStop}%
\bibitem [{\citenamefont {Chen}\ \emph {et~al.}(2023)\citenamefont {Chen},
  \citenamefont {Wang}, \citenamefont {Yu}, \citenamefont {Li},\ and\
  \citenamefont {Guo}}]{25chen2023continuous}%
  \BibitemOpen
  \bibfield  {author} {\bibinfo {author} {\bibfnamefont {Z.}~\bibnamefont
  {Chen}}, \bibinfo {author} {\bibfnamefont {X.}~\bibnamefont {Wang}}, \bibinfo
  {author} {\bibfnamefont {S.}~\bibnamefont {Yu}}, \bibinfo {author}
  {\bibfnamefont {Z.}~\bibnamefont {Li}},\ and\ \bibinfo {author}
  {\bibfnamefont {H.}~\bibnamefont {Guo}},\ }\bibinfo {title} {Continuous-mode
  quantum key distribution with digital signal processing},\ \href@noop {}
  {\bibfield  {journal} {\bibinfo  {journal} {npj Quantum Inform.}\ }\textbf
  {\bibinfo {volume} {9}},\ \bibinfo {pages} {28} (\bibinfo {year}
  {2023})}\BibitemShut {NoStop}%
\bibitem [{\citenamefont {Zhang}\ \emph {et~al.}(2023)\citenamefont {Zhang},
  \citenamefont {Liu}, \citenamefont {Qi}, \citenamefont {He},\ and\
  \citenamefont {Huang}}]{zhang2023automatic}%
  \BibitemOpen
  \bibfield  {author} {\bibinfo {author} {\bibfnamefont {Z.-K.}\ \bibnamefont
  {Zhang}}, \bibinfo {author} {\bibfnamefont {W.-Q.}\ \bibnamefont {Liu}},
  \bibinfo {author} {\bibfnamefont {J.}~\bibnamefont {Qi}}, \bibinfo {author}
  {\bibfnamefont {C.}~\bibnamefont {He}},\ and\ \bibinfo {author}
  {\bibfnamefont {P.}~\bibnamefont {Huang}},\ }\bibinfo {title} {Automatic
  phase compensation of a continuous-variable quantum-key-distribution system
  via deep learning},\ \href@noop {} {\bibfield  {journal} {\bibinfo  {journal}
  {Phys. Rev. A}\ }\textbf {\bibinfo {volume} {107}},\ \bibinfo {pages}
  {062614} (\bibinfo {year} {2023})}\BibitemShut {NoStop}%
\bibitem [{\citenamefont {Kanitschar}\ \emph {et~al.}(2023)\citenamefont
  {Kanitschar}, \citenamefont {George}, \citenamefont {Lin}, \citenamefont
  {Upadhyaya},\ and\ \citenamefont {L{\"u}tkenhaus}}]{kanitschar2023finite}%
  \BibitemOpen
  \bibfield  {author} {\bibinfo {author} {\bibfnamefont {F.}~\bibnamefont
  {Kanitschar}}, \bibinfo {author} {\bibfnamefont {I.}~\bibnamefont {George}},
  \bibinfo {author} {\bibfnamefont {J.}~\bibnamefont {Lin}}, \bibinfo {author}
  {\bibfnamefont {T.}~\bibnamefont {Upadhyaya}},\ and\ \bibinfo {author}
  {\bibfnamefont {N.}~\bibnamefont {L{\"u}tkenhaus}},\ }\bibinfo {title}
  {Finite-size security for discrete-modulated continuous-variable quantum key
  distribution protocols},\ \href@noop {} {\bibfield  {journal} {\bibinfo
  {journal} {PRX Quantum}\ }\textbf {\bibinfo {volume} {4}},\ \bibinfo {pages}
  {040306} (\bibinfo {year} {2023})}\BibitemShut {NoStop}%
\bibitem [{\citenamefont {Primaatmaja}\ \emph {et~al.}(2024)\citenamefont
  {Primaatmaja}, \citenamefont {Kon},\ and\ \citenamefont
  {Lim}}]{primaatmaja2024discrete}%
  \BibitemOpen
  \bibfield  {author} {\bibinfo {author} {\bibfnamefont {I.~W.}\ \bibnamefont
  {Primaatmaja}}, \bibinfo {author} {\bibfnamefont {W.~Y.}\ \bibnamefont
  {Kon}},\ and\ \bibinfo {author} {\bibfnamefont {C.}~\bibnamefont {Lim}},\
  }\bibinfo {title} {Discrete-modulated continuous-variable quantum key
  distribution secure against general attacks},\ \href@noop {} {\bibfield
  {journal} {\bibinfo  {journal} {arXiv preprint arXiv:2409.02630}\ } (\bibinfo
  {year} {2024})}\BibitemShut {NoStop}%
\bibitem [{\citenamefont {Jouguet}\ \emph {et~al.}(2013)\citenamefont
  {Jouguet}, \citenamefont {Kunz-Jacques}, \citenamefont {Leverrier},
  \citenamefont {Grangier},\ and\ \citenamefont
  {Diamanti}}]{13jouguet2013experimental}%
  \BibitemOpen
  \bibfield  {author} {\bibinfo {author} {\bibfnamefont {P.}~\bibnamefont
  {Jouguet}}, \bibinfo {author} {\bibfnamefont {S.}~\bibnamefont
  {Kunz-Jacques}}, \bibinfo {author} {\bibfnamefont {A.}~\bibnamefont
  {Leverrier}}, \bibinfo {author} {\bibfnamefont {P.}~\bibnamefont
  {Grangier}},\ and\ \bibinfo {author} {\bibfnamefont {E.}~\bibnamefont
  {Diamanti}},\ }\bibinfo {title} {Experimental demonstration of long-distance
  continuous-variable quantum key distribution},\ \href@noop {} {\bibfield
  {journal} {\bibinfo  {journal} {Nat. Photonics}\ }\textbf {\bibinfo {volume}
  {7}},\ \bibinfo {pages} {378} (\bibinfo {year} {2013})}\BibitemShut {NoStop}%
\bibitem [{\citenamefont {Qi}\ \emph {et~al.}(2015)\citenamefont {Qi},
  \citenamefont {Lougovski}, \citenamefont {Pooser}, \citenamefont {Grice},\
  and\ \citenamefont {Bobrek}}]{qi2015generating}%
  \BibitemOpen
  \bibfield  {author} {\bibinfo {author} {\bibfnamefont {B.}~\bibnamefont
  {Qi}}, \bibinfo {author} {\bibfnamefont {P.}~\bibnamefont {Lougovski}},
  \bibinfo {author} {\bibfnamefont {R.}~\bibnamefont {Pooser}}, \bibinfo
  {author} {\bibfnamefont {W.}~\bibnamefont {Grice}},\ and\ \bibinfo {author}
  {\bibfnamefont {M.}~\bibnamefont {Bobrek}},\ }\bibinfo {title} {Generating
  the local oscillator “locally” in continuous-variable quantum key
  distribution based on coherent detection},\ \href@noop {} {\bibfield
  {journal} {\bibinfo  {journal} {Phys. Rev. X}\ }\textbf {\bibinfo {volume}
  {5}},\ \bibinfo {pages} {041009} (\bibinfo {year} {2015})}\BibitemShut
  {NoStop}%
\bibitem [{\citenamefont {Soh}\ \emph {et~al.}(2015)\citenamefont {Soh},
  \citenamefont {Brif}, \citenamefont {Coles}, \citenamefont {L{\"u}tkenhaus},
  \citenamefont {Camacho}, \citenamefont {Urayama},\ and\ \citenamefont
  {Sarovar}}]{soh2015self}%
  \BibitemOpen
  \bibfield  {author} {\bibinfo {author} {\bibfnamefont {D.~B.}\ \bibnamefont
  {Soh}}, \bibinfo {author} {\bibfnamefont {C.}~\bibnamefont {Brif}}, \bibinfo
  {author} {\bibfnamefont {P.~J.}\ \bibnamefont {Coles}}, \bibinfo {author}
  {\bibfnamefont {N.}~\bibnamefont {L{\"u}tkenhaus}}, \bibinfo {author}
  {\bibfnamefont {R.~M.}\ \bibnamefont {Camacho}}, \bibinfo {author}
  {\bibfnamefont {J.}~\bibnamefont {Urayama}},\ and\ \bibinfo {author}
  {\bibfnamefont {M.}~\bibnamefont {Sarovar}},\ }\bibinfo {title}
  {Self-referenced continuous-variable quantum key distribution protocol},\
  \href@noop {} {\bibfield  {journal} {\bibinfo  {journal} {Phys. Rev. X}\
  }\textbf {\bibinfo {volume} {5}},\ \bibinfo {pages} {041010} (\bibinfo {year}
  {2015})}\BibitemShut {NoStop}%
\bibitem [{\citenamefont {Huang}\ \emph {et~al.}(2016)\citenamefont {Huang},
  \citenamefont {Huang}, \citenamefont {Lin},\ and\ \citenamefont
  {Zeng}}]{14huang2016long}%
  \BibitemOpen
  \bibfield  {author} {\bibinfo {author} {\bibfnamefont {D.}~\bibnamefont
  {Huang}}, \bibinfo {author} {\bibfnamefont {P.}~\bibnamefont {Huang}},
  \bibinfo {author} {\bibfnamefont {D.}~\bibnamefont {Lin}},\ and\ \bibinfo
  {author} {\bibfnamefont {G.}~\bibnamefont {Zeng}},\ }\bibinfo {title}
  {Long-distance continuous-variable quantum key distribution by controlling
  excess noise},\ \href@noop {} {\bibfield  {journal} {\bibinfo  {journal}
  {Sci. Rep.}\ }\textbf {\bibinfo {volume} {6}},\ \bibinfo {pages} {19201}
  (\bibinfo {year} {2016})}\BibitemShut {NoStop}%
\bibitem [{\citenamefont {Hajomer}\ \emph
  {et~al.}(2024{\natexlab{a}})\citenamefont {Hajomer}, \citenamefont {Derkach},
  \citenamefont {Jain}, \citenamefont {Chin}, \citenamefont {Andersen},\ and\
  \citenamefont {Gehring}}]{16hajomer2024long}%
  \BibitemOpen
  \bibfield  {author} {\bibinfo {author} {\bibfnamefont {A.~A.}\ \bibnamefont
  {Hajomer}}, \bibinfo {author} {\bibfnamefont {I.}~\bibnamefont {Derkach}},
  \bibinfo {author} {\bibfnamefont {N.}~\bibnamefont {Jain}}, \bibinfo {author}
  {\bibfnamefont {H.-M.}\ \bibnamefont {Chin}}, \bibinfo {author}
  {\bibfnamefont {U.~L.}\ \bibnamefont {Andersen}},\ and\ \bibinfo {author}
  {\bibfnamefont {T.}~\bibnamefont {Gehring}},\ }\bibinfo {title}
  {Long-distance continuous-variable quantum key distribution over 100-km fiber
  with local local oscillator},\ \href@noop {} {\bibfield  {journal} {\bibinfo
  {journal} {Sci. Adv.}\ }\textbf {\bibinfo {volume} {10}},\ \bibinfo {pages}
  {eadi9474} (\bibinfo {year} {2024}{\natexlab{a}})}\BibitemShut {NoStop}%
\bibitem [{\citenamefont {Williams}\ \emph {et~al.}(2024)\citenamefont
  {Williams}, \citenamefont {Qi}, \citenamefont {Alshowkan}, \citenamefont
  {Evans},\ and\ \citenamefont {Peters}}]{williams2024field}%
  \BibitemOpen
  \bibfield  {author} {\bibinfo {author} {\bibfnamefont {B.~P.}\ \bibnamefont
  {Williams}}, \bibinfo {author} {\bibfnamefont {B.}~\bibnamefont {Qi}},
  \bibinfo {author} {\bibfnamefont {M.}~\bibnamefont {Alshowkan}}, \bibinfo
  {author} {\bibfnamefont {P.~G.}\ \bibnamefont {Evans}},\ and\ \bibinfo
  {author} {\bibfnamefont {N.~A.}\ \bibnamefont {Peters}},\ }\bibinfo {title}
  {Field test of continuous-variable quantum key distribution with a true local
  oscillator},\ \href@noop {} {\bibfield  {journal} {\bibinfo  {journal} {Phys.
  Rev. Appl.}\ }\textbf {\bibinfo {volume} {21}},\ \bibinfo {pages} {014056}
  (\bibinfo {year} {2024})}\BibitemShut {NoStop}%
\bibitem [{\citenamefont {Wang}\ \emph
  {et~al.}(2022{\natexlab{a}})\citenamefont {Wang}, \citenamefont {Li},
  \citenamefont {Pi}, \citenamefont {Pan}, \citenamefont {Shao}, \citenamefont
  {Ma}, \citenamefont {Zhang}, \citenamefont {Yang}, \citenamefont {Zhang},
  \citenamefont {Huang} \emph {et~al.}}]{15wang2022sub}%
  \BibitemOpen
  \bibfield  {author} {\bibinfo {author} {\bibfnamefont {H.}~\bibnamefont
  {Wang}}, \bibinfo {author} {\bibfnamefont {Y.}~\bibnamefont {Li}}, \bibinfo
  {author} {\bibfnamefont {Y.}~\bibnamefont {Pi}}, \bibinfo {author}
  {\bibfnamefont {Y.}~\bibnamefont {Pan}}, \bibinfo {author} {\bibfnamefont
  {Y.}~\bibnamefont {Shao}}, \bibinfo {author} {\bibfnamefont {L.}~\bibnamefont
  {Ma}}, \bibinfo {author} {\bibfnamefont {Y.}~\bibnamefont {Zhang}}, \bibinfo
  {author} {\bibfnamefont {J.}~\bibnamefont {Yang}}, \bibinfo {author}
  {\bibfnamefont {T.}~\bibnamefont {Zhang}}, \bibinfo {author} {\bibfnamefont
  {W.}~\bibnamefont {Huang}}, \emph {et~al.},\ }\bibinfo {title} {Sub-gbps key
  rate four-state continuous-variable quantum key distribution within
  metropolitan area},\ \href@noop {} {\bibfield  {journal} {\bibinfo  {journal}
  {Commun. Phys.}\ }\textbf {\bibinfo {volume} {5}},\ \bibinfo {pages} {162}
  (\bibinfo {year} {2022}{\natexlab{a}})}\BibitemShut {NoStop}%
\bibitem [{\citenamefont {Tian}\ \emph {et~al.}(2023)\citenamefont {Tian},
  \citenamefont {Zhang}, \citenamefont {Liu}, \citenamefont {Wang},
  \citenamefont {Lu}, \citenamefont {Wang},\ and\ \citenamefont
  {Li}}]{tian2023high}%
  \BibitemOpen
  \bibfield  {author} {\bibinfo {author} {\bibfnamefont {Y.}~\bibnamefont
  {Tian}}, \bibinfo {author} {\bibfnamefont {Y.}~\bibnamefont {Zhang}},
  \bibinfo {author} {\bibfnamefont {S.}~\bibnamefont {Liu}}, \bibinfo {author}
  {\bibfnamefont {P.}~\bibnamefont {Wang}}, \bibinfo {author} {\bibfnamefont
  {Z.}~\bibnamefont {Lu}}, \bibinfo {author} {\bibfnamefont {X.}~\bibnamefont
  {Wang}},\ and\ \bibinfo {author} {\bibfnamefont {Y.}~\bibnamefont {Li}},\
  }\bibinfo {title} {High-performance long-distance discrete-modulation
  continuous-variable quantum key distribution},\ \href@noop {} {\bibfield
  {journal} {\bibinfo  {journal} {Opt. Lett.}\ }\textbf {\bibinfo {volume}
  {48}},\ \bibinfo {pages} {2953} (\bibinfo {year} {2023})}\BibitemShut
  {NoStop}%
\bibitem [{\citenamefont {Xu}\ \emph {et~al.}(2023{\natexlab{a}})\citenamefont
  {Xu}, \citenamefont {Wang}, \citenamefont {Li}, \citenamefont {Zhao},
  \citenamefont {Huang},\ and\ \citenamefont {Zeng}}]{xu2023simultaneous}%
  \BibitemOpen
  \bibfield  {author} {\bibinfo {author} {\bibfnamefont {Y.}~\bibnamefont
  {Xu}}, \bibinfo {author} {\bibfnamefont {T.}~\bibnamefont {Wang}}, \bibinfo
  {author} {\bibfnamefont {L.}~\bibnamefont {Li}}, \bibinfo {author}
  {\bibfnamefont {H.}~\bibnamefont {Zhao}}, \bibinfo {author} {\bibfnamefont
  {P.}~\bibnamefont {Huang}},\ and\ \bibinfo {author} {\bibfnamefont
  {G.}~\bibnamefont {Zeng}},\ }\bibinfo {title} {Simultaneous
  continuous-variable quantum key distribution and classical optical
  communication over a shared infrastructure},\ \href@noop {} {\bibfield
  {journal} {\bibinfo  {journal} {Appl. Phys. Lett.}\ }\textbf {\bibinfo
  {volume} {123}} (\bibinfo {year} {2023}{\natexlab{a}})}\BibitemShut {NoStop}%
\bibitem [{\citenamefont {Jaksch}\ \emph {et~al.}(2024)\citenamefont {Jaksch},
  \citenamefont {Dirmeier}, \citenamefont {Weiser}, \citenamefont {Richter},
  \citenamefont {Bayraktar}, \citenamefont {Hacker}, \citenamefont
  {R{\"o}sler}, \citenamefont {Khan}, \citenamefont {Petscharning},
  \citenamefont {Grafenauer} \emph {et~al.}}]{jaksch2024composable}%
  \BibitemOpen
  \bibfield  {author} {\bibinfo {author} {\bibfnamefont {K.}~\bibnamefont
  {Jaksch}}, \bibinfo {author} {\bibfnamefont {T.}~\bibnamefont {Dirmeier}},
  \bibinfo {author} {\bibfnamefont {Y.}~\bibnamefont {Weiser}}, \bibinfo
  {author} {\bibfnamefont {S.}~\bibnamefont {Richter}}, \bibinfo {author}
  {\bibfnamefont {{\"O}.}~\bibnamefont {Bayraktar}}, \bibinfo {author}
  {\bibfnamefont {B.}~\bibnamefont {Hacker}}, \bibinfo {author} {\bibfnamefont
  {C.}~\bibnamefont {R{\"o}sler}}, \bibinfo {author} {\bibfnamefont
  {I.}~\bibnamefont {Khan}}, \bibinfo {author} {\bibfnamefont {S.}~\bibnamefont
  {Petscharning}}, \bibinfo {author} {\bibfnamefont {T.}~\bibnamefont
  {Grafenauer}}, \emph {et~al.},\ }\bibinfo {title} {Composable free-space
  continuous-variable quantum key distribution using discrete modulation},\
  \href@noop {} {\bibfield  {journal} {\bibinfo  {journal} {arXiv preprint
  arXiv:2410.12915}\ } (\bibinfo {year} {2024})}\BibitemShut {NoStop}%
\bibitem [{\citenamefont {Hajomer}\ \emph
  {et~al.}(2024{\natexlab{b}})\citenamefont {Hajomer}, \citenamefont
  {Kanitschar}, \citenamefont {Jain}, \citenamefont {Hentschel}, \citenamefont
  {Zhang}, \citenamefont {L{\"u}tkenhaus}, \citenamefont {Andersen},
  \citenamefont {Pacher},\ and\ \citenamefont
  {Gehring}}]{hajomer2024experimental}%
  \BibitemOpen
  \bibfield  {author} {\bibinfo {author} {\bibfnamefont {A.~A.}\ \bibnamefont
  {Hajomer}}, \bibinfo {author} {\bibfnamefont {F.}~\bibnamefont {Kanitschar}},
  \bibinfo {author} {\bibfnamefont {N.}~\bibnamefont {Jain}}, \bibinfo {author}
  {\bibfnamefont {M.}~\bibnamefont {Hentschel}}, \bibinfo {author}
  {\bibfnamefont {R.}~\bibnamefont {Zhang}}, \bibinfo {author} {\bibfnamefont
  {N.}~\bibnamefont {L{\"u}tkenhaus}}, \bibinfo {author} {\bibfnamefont
  {U.~L.}\ \bibnamefont {Andersen}}, \bibinfo {author} {\bibfnamefont
  {C.}~\bibnamefont {Pacher}},\ and\ \bibinfo {author} {\bibfnamefont
  {T.}~\bibnamefont {Gehring}},\ }\bibinfo {title} {Experimental composable key
  distribution using discrete-modulated continuous variable quantum
  cryptography},\ \href@noop {} {\bibfield  {journal} {\bibinfo  {journal}
  {arXiv preprint arXiv:2410.13702}\ } (\bibinfo {year}
  {2024}{\natexlab{b}})}\BibitemShut {NoStop}%
\bibitem [{\citenamefont {Wang}\ \emph
  {et~al.}(2022{\natexlab{b}})\citenamefont {Wang}, \citenamefont {Wang},
  \citenamefont {Zhou}, \citenamefont {Chen}, \citenamefont {Yu},\ and\
  \citenamefont {Guo}}]{17post_wang2022continuous}%
  \BibitemOpen
  \bibfield  {author} {\bibinfo {author} {\bibfnamefont {X.}~\bibnamefont
  {Wang}}, \bibinfo {author} {\bibfnamefont {H.}~\bibnamefont {Wang}}, \bibinfo
  {author} {\bibfnamefont {C.}~\bibnamefont {Zhou}}, \bibinfo {author}
  {\bibfnamefont {Z.}~\bibnamefont {Chen}}, \bibinfo {author} {\bibfnamefont
  {S.}~\bibnamefont {Yu}},\ and\ \bibinfo {author} {\bibfnamefont
  {H.}~\bibnamefont {Guo}},\ }\bibinfo {title} {Continuous-variable quantum key
  distribution with low-complexity information reconciliation},\ \href@noop {}
  {\bibfield  {journal} {\bibinfo  {journal} {Opt. Express}\ }\textbf {\bibinfo
  {volume} {30}},\ \bibinfo {pages} {30455} (\bibinfo {year}
  {2022}{\natexlab{b}})}\BibitemShut {NoStop}%
\bibitem [{\citenamefont {Cao}\ \emph {et~al.}(2023)\citenamefont {Cao},
  \citenamefont {Chen}, \citenamefont {Chai}, \citenamefont {Liang},\ and\
  \citenamefont {Yuan}}]{18post_cao2023rate}%
  \BibitemOpen
  \bibfield  {author} {\bibinfo {author} {\bibfnamefont {Z.}~\bibnamefont
  {Cao}}, \bibinfo {author} {\bibfnamefont {X.}~\bibnamefont {Chen}}, \bibinfo
  {author} {\bibfnamefont {G.}~\bibnamefont {Chai}}, \bibinfo {author}
  {\bibfnamefont {K.}~\bibnamefont {Liang}},\ and\ \bibinfo {author}
  {\bibfnamefont {Y.}~\bibnamefont {Yuan}},\ }\bibinfo {title} {Rate-adaptive
  polar-coding-based reconciliation for continuous-variable quantum key
  distribution at low signal-to-noise ratio},\ \href@noop {} {\bibfield
  {journal} {\bibinfo  {journal} {Phys. Rev. Appl.}\ }\textbf {\bibinfo
  {volume} {19}},\ \bibinfo {pages} {044023} (\bibinfo {year}
  {2023})}\BibitemShut {NoStop}%
\bibitem [{\citenamefont {Wang}\ \emph
  {et~al.}(2023{\natexlab{a}})\citenamefont {Wang}, \citenamefont {Xu},
  \citenamefont {Zhao}, \citenamefont {Chen}, \citenamefont {Yu},\ and\
  \citenamefont {Guo}}]{19post_wang2023non}%
  \BibitemOpen
  \bibfield  {author} {\bibinfo {author} {\bibfnamefont {X.}~\bibnamefont
  {Wang}}, \bibinfo {author} {\bibfnamefont {M.}~\bibnamefont {Xu}}, \bibinfo
  {author} {\bibfnamefont {Y.}~\bibnamefont {Zhao}}, \bibinfo {author}
  {\bibfnamefont {Z.}~\bibnamefont {Chen}}, \bibinfo {author} {\bibfnamefont
  {S.}~\bibnamefont {Yu}},\ and\ \bibinfo {author} {\bibfnamefont
  {H.}~\bibnamefont {Guo}},\ }\bibinfo {title} {Non-gaussian reconciliation for
  continuous-variable quantum key distribution},\ \href@noop {} {\bibfield
  {journal} {\bibinfo  {journal} {Phys. Rev. Appl.}\ }\textbf {\bibinfo
  {volume} {19}},\ \bibinfo {pages} {054084} (\bibinfo {year}
  {2023}{\natexlab{a}})}\BibitemShut {NoStop}%
\bibitem [{\citenamefont {Wang}\ \emph
  {et~al.}(2023{\natexlab{b}})\citenamefont {Wang}, \citenamefont {Chen},
  \citenamefont {Li}, \citenamefont {Qi}, \citenamefont {Yu},\ and\
  \citenamefont {Guo}}]{20network_wang2023experimental}%
  \BibitemOpen
  \bibfield  {author} {\bibinfo {author} {\bibfnamefont {X.}~\bibnamefont
  {Wang}}, \bibinfo {author} {\bibfnamefont {Z.}~\bibnamefont {Chen}}, \bibinfo
  {author} {\bibfnamefont {Z.}~\bibnamefont {Li}}, \bibinfo {author}
  {\bibfnamefont {D.}~\bibnamefont {Qi}}, \bibinfo {author} {\bibfnamefont
  {S.}~\bibnamefont {Yu}},\ and\ \bibinfo {author} {\bibfnamefont
  {H.}~\bibnamefont {Guo}},\ }\bibinfo {title} {Experimental upstream
  transmission of continuous variable quantum key distribution access
  network},\ \href@noop {} {\bibfield  {journal} {\bibinfo  {journal} {Opt.
  Lett.}\ }\textbf {\bibinfo {volume} {48}},\ \bibinfo {pages} {3327} (\bibinfo
  {year} {2023}{\natexlab{b}})}\BibitemShut {NoStop}%
\bibitem [{\citenamefont {Xu}\ \emph {et~al.}(2023{\natexlab{b}})\citenamefont
  {Xu}, \citenamefont {Wang}, \citenamefont {Zhao}, \citenamefont {Huang},\
  and\ \citenamefont {Zeng}}]{21network_xu2023round}%
  \BibitemOpen
  \bibfield  {author} {\bibinfo {author} {\bibfnamefont {Y.}~\bibnamefont
  {Xu}}, \bibinfo {author} {\bibfnamefont {T.}~\bibnamefont {Wang}}, \bibinfo
  {author} {\bibfnamefont {H.}~\bibnamefont {Zhao}}, \bibinfo {author}
  {\bibfnamefont {P.}~\bibnamefont {Huang}},\ and\ \bibinfo {author}
  {\bibfnamefont {G.}~\bibnamefont {Zeng}},\ }\bibinfo {title} {Round-trip
  multi-band quantum access network},\ \href@noop {} {\bibfield  {journal}
  {\bibinfo  {journal} {Photonics Res.}\ }\textbf {\bibinfo {volume} {11}},\
  \bibinfo {pages} {1449} (\bibinfo {year} {2023}{\natexlab{b}})}\BibitemShut
  {NoStop}%
\bibitem [{\citenamefont {Qi}\ \emph {et~al.}(2024)\citenamefont {Qi},
  \citenamefont {Wang}, \citenamefont {Li}, \citenamefont {Ma}, \citenamefont
  {Chen}, \citenamefont {Lu},\ and\ \citenamefont
  {Yu}}]{22network_qi2024experimental}%
  \BibitemOpen
  \bibfield  {author} {\bibinfo {author} {\bibfnamefont {D.}~\bibnamefont
  {Qi}}, \bibinfo {author} {\bibfnamefont {X.}~\bibnamefont {Wang}}, \bibinfo
  {author} {\bibfnamefont {Z.}~\bibnamefont {Li}}, \bibinfo {author}
  {\bibfnamefont {J.}~\bibnamefont {Ma}}, \bibinfo {author} {\bibfnamefont
  {Z.}~\bibnamefont {Chen}}, \bibinfo {author} {\bibfnamefont {Y.}~\bibnamefont
  {Lu}},\ and\ \bibinfo {author} {\bibfnamefont {S.}~\bibnamefont {Yu}},\
  }\bibinfo {title} {Experimental demonstration of a quantum downstream access
  network in continuous variable quantum key distribution with a local local
  oscillator},\ \href@noop {} {\bibfield  {journal} {\bibinfo  {journal}
  {Photonics Res.}\ }\textbf {\bibinfo {volume} {12}},\ \bibinfo {pages} {1262}
  (\bibinfo {year} {2024})}\BibitemShut {NoStop}%
\bibitem [{\citenamefont {Li}\ \emph {et~al.}(2024)\citenamefont {Li},
  \citenamefont {Wang}, \citenamefont {Qi}, \citenamefont {Chen},\ and\
  \citenamefont {Yu}}]{23network_li2024experimental}%
  \BibitemOpen
  \bibfield  {author} {\bibinfo {author} {\bibfnamefont {Z.}~\bibnamefont
  {Li}}, \bibinfo {author} {\bibfnamefont {X.}~\bibnamefont {Wang}}, \bibinfo
  {author} {\bibfnamefont {D.}~\bibnamefont {Qi}}, \bibinfo {author}
  {\bibfnamefont {Z.}~\bibnamefont {Chen}},\ and\ \bibinfo {author}
  {\bibfnamefont {S.}~\bibnamefont {Yu}},\ }\bibinfo {title} {Experimental
  implementation of four-user downstream access network continuous-variable
  quantum key distribution},\ \href@noop {} {\bibfield  {journal} {\bibinfo
  {journal} {J. Lightwave Technol.}\ } (\bibinfo {year} {2024})}\BibitemShut
  {NoStop}%
\bibitem [{\citenamefont {Bian}\ \emph {et~al.}(2023)\citenamefont {Bian},
  \citenamefont {Zhang}, \citenamefont {Zhou}, \citenamefont {Yu},
  \citenamefont {Li},\ and\ \citenamefont {Guo}}]{bian2023high}%
  \BibitemOpen
  \bibfield  {author} {\bibinfo {author} {\bibfnamefont {Y.}~\bibnamefont
  {Bian}}, \bibinfo {author} {\bibfnamefont {Y.-C.}\ \bibnamefont {Zhang}},
  \bibinfo {author} {\bibfnamefont {C.}~\bibnamefont {Zhou}}, \bibinfo {author}
  {\bibfnamefont {S.}~\bibnamefont {Yu}}, \bibinfo {author} {\bibfnamefont
  {Z.}~\bibnamefont {Li}},\ and\ \bibinfo {author} {\bibfnamefont
  {H.}~\bibnamefont {Guo}},\ }\bibinfo {title} {High-rate point-to-multipoint
  quantum key distribution using coherent states},\ \href@noop {} {\bibfield
  {journal} {\bibinfo  {journal} {arXiv preprint arXiv:2302.02391}\ } (\bibinfo
  {year} {2023})}\BibitemShut {NoStop}%
\bibitem [{\citenamefont {Hajomer}\ \emph
  {et~al.}(2024{\natexlab{c}})\citenamefont {Hajomer}, \citenamefont {Derkach},
  \citenamefont {Filip}, \citenamefont {Andersen}, \citenamefont {C.~Usenko},\
  and\ \citenamefont {Gehring}}]{hajomer2024continuous}%
  \BibitemOpen
  \bibfield  {author} {\bibinfo {author} {\bibfnamefont {A.~A.}\ \bibnamefont
  {Hajomer}}, \bibinfo {author} {\bibfnamefont {I.}~\bibnamefont {Derkach}},
  \bibinfo {author} {\bibfnamefont {R.}~\bibnamefont {Filip}}, \bibinfo
  {author} {\bibfnamefont {U.~L.}\ \bibnamefont {Andersen}}, \bibinfo {author}
  {\bibfnamefont {V.}~\bibnamefont {C.~Usenko}},\ and\ \bibinfo {author}
  {\bibfnamefont {T.}~\bibnamefont {Gehring}},\ }\bibinfo {title}
  {Continuous-variable quantum passive optical network},\ \href@noop {}
  {\bibfield  {journal} {\bibinfo  {journal} {Light Sci. Appl.}\ }\textbf
  {\bibinfo {volume} {13}},\ \bibinfo {pages} {291} (\bibinfo {year}
  {2024}{\natexlab{c}})}\BibitemShut {NoStop}%
\bibitem [{\citenamefont {Kanitschar}\ and\ \citenamefont
  {Pacher}(2024)}]{kanitschar2024security}%
  \BibitemOpen
  \bibfield  {author} {\bibinfo {author} {\bibfnamefont {F.}~\bibnamefont
  {Kanitschar}}\ and\ \bibinfo {author} {\bibfnamefont {C.}~\bibnamefont
  {Pacher}},\ }\bibinfo {title} {Security of multi-user quantum key
  distribution with discrete-modulated continuous-variables},\ \href@noop {}
  {\bibfield  {journal} {\bibinfo  {journal} {arXiv preprint arXiv:2406.14610}\
  } (\bibinfo {year} {2024})}\BibitemShut {NoStop}%
\bibitem [{\citenamefont {Liu}\ \emph {et~al.}(2022)\citenamefont {Liu},
  \citenamefont {Cao}, \citenamefont {Wang}, \citenamefont {Liu}, \citenamefont
  {Lu}, \citenamefont {Wang},\ and\ \citenamefont
  {Li}}]{24specbroad_liu2022impact}%
  \BibitemOpen
  \bibfield  {author} {\bibinfo {author} {\bibfnamefont {J.}~\bibnamefont
  {Liu}}, \bibinfo {author} {\bibfnamefont {Y.}~\bibnamefont {Cao}}, \bibinfo
  {author} {\bibfnamefont {P.}~\bibnamefont {Wang}}, \bibinfo {author}
  {\bibfnamefont {S.}~\bibnamefont {Liu}}, \bibinfo {author} {\bibfnamefont
  {Z.}~\bibnamefont {Lu}}, \bibinfo {author} {\bibfnamefont {X.}~\bibnamefont
  {Wang}},\ and\ \bibinfo {author} {\bibfnamefont {Y.}~\bibnamefont {Li}},\
  }\bibinfo {title} {Impact of homodyne receiver bandwidth and signal
  modulation patterns on the continuous-variable quantum key distribution},\
  \href@noop {} {\bibfield  {journal} {\bibinfo  {journal} {Opt. Express}\
  }\textbf {\bibinfo {volume} {30}},\ \bibinfo {pages} {27912} (\bibinfo {year}
  {2022})}\BibitemShut {NoStop}%
\bibitem [{\citenamefont {Li}\ \emph {et~al.}(2014)\citenamefont {Li},
  \citenamefont {Zhang}, \citenamefont {Xu}, \citenamefont {Peng},\ and\
  \citenamefont {Guo}}]{26MDI_li2014continuous}%
  \BibitemOpen
  \bibfield  {author} {\bibinfo {author} {\bibfnamefont {Z.}~\bibnamefont
  {Li}}, \bibinfo {author} {\bibfnamefont {Y.}~\bibnamefont {Zhang}}, \bibinfo
  {author} {\bibfnamefont {F.}~\bibnamefont {Xu}}, \bibinfo {author}
  {\bibfnamefont {X.}~\bibnamefont {Peng}},\ and\ \bibinfo {author}
  {\bibfnamefont {H.}~\bibnamefont {Guo}},\ }\bibinfo {title}
  {Continuous-variable measurement-device-independent quantum key
  distribution},\ \href@noop {} {\bibfield  {journal} {\bibinfo  {journal}
  {Phys. Rev. A}\ }\textbf {\bibinfo {volume} {89}},\ \bibinfo {pages} {052301}
  (\bibinfo {year} {2014})}\BibitemShut {NoStop}%
\bibitem [{\citenamefont {Pirandola}\ \emph {et~al.}(2015)\citenamefont
  {Pirandola}, \citenamefont {Ottaviani}, \citenamefont {Spedalieri},
  \citenamefont {Weedbrook}, \citenamefont {Braunstein}, \citenamefont {Lloyd},
  \citenamefont {Gehring}, \citenamefont {Jacobsen},\ and\ \citenamefont
  {Andersen}}]{27MDI_pirandola2015high}%
  \BibitemOpen
  \bibfield  {author} {\bibinfo {author} {\bibfnamefont {S.}~\bibnamefont
  {Pirandola}}, \bibinfo {author} {\bibfnamefont {C.}~\bibnamefont
  {Ottaviani}}, \bibinfo {author} {\bibfnamefont {G.}~\bibnamefont
  {Spedalieri}}, \bibinfo {author} {\bibfnamefont {C.}~\bibnamefont
  {Weedbrook}}, \bibinfo {author} {\bibfnamefont {S.~L.}\ \bibnamefont
  {Braunstein}}, \bibinfo {author} {\bibfnamefont {S.}~\bibnamefont {Lloyd}},
  \bibinfo {author} {\bibfnamefont {T.}~\bibnamefont {Gehring}}, \bibinfo
  {author} {\bibfnamefont {C.~S.}\ \bibnamefont {Jacobsen}},\ and\ \bibinfo
  {author} {\bibfnamefont {U.~L.}\ \bibnamefont {Andersen}},\ }\bibinfo {title}
  {High-rate measurement-device-independent quantum cryptography},\ \href@noop
  {} {\bibfield  {journal} {\bibinfo  {journal} {Nat. Photonics}\ }\textbf
  {\bibinfo {volume} {9}},\ \bibinfo {pages} {397} (\bibinfo {year}
  {2015})}\BibitemShut {NoStop}%
\bibitem [{\citenamefont {Lupo}\ \emph
  {et~al.}(2018{\natexlab{a}})\citenamefont {Lupo}, \citenamefont {Ottaviani},
  \citenamefont {Papanastasiou},\ and\ \citenamefont
  {Pirandola}}]{28MDI_lupo2018parameter}%
  \BibitemOpen
  \bibfield  {author} {\bibinfo {author} {\bibfnamefont {C.}~\bibnamefont
  {Lupo}}, \bibinfo {author} {\bibfnamefont {C.}~\bibnamefont {Ottaviani}},
  \bibinfo {author} {\bibfnamefont {P.}~\bibnamefont {Papanastasiou}},\ and\
  \bibinfo {author} {\bibfnamefont {S.}~\bibnamefont {Pirandola}},\ }\bibinfo
  {title} {Parameter estimation with almost no public communication for
  continuous-variable quantum key distribution},\ \href@noop {} {\bibfield
  {journal} {\bibinfo  {journal} {Phys. Rev. Lett.}\ }\textbf {\bibinfo
  {volume} {120}},\ \bibinfo {pages} {220505} (\bibinfo {year}
  {2018}{\natexlab{a}})}\BibitemShut {NoStop}%
\bibitem [{\citenamefont {Tian}\ \emph {et~al.}(2022)\citenamefont {Tian},
  \citenamefont {Wang}, \citenamefont {Liu}, \citenamefont {Du}, \citenamefont
  {Liu}, \citenamefont {Lu}, \citenamefont {Wang},\ and\ \citenamefont
  {Li}}]{29MDI_tian2022experimental}%
  \BibitemOpen
  \bibfield  {author} {\bibinfo {author} {\bibfnamefont {Y.}~\bibnamefont
  {Tian}}, \bibinfo {author} {\bibfnamefont {P.}~\bibnamefont {Wang}}, \bibinfo
  {author} {\bibfnamefont {J.}~\bibnamefont {Liu}}, \bibinfo {author}
  {\bibfnamefont {S.}~\bibnamefont {Du}}, \bibinfo {author} {\bibfnamefont
  {W.}~\bibnamefont {Liu}}, \bibinfo {author} {\bibfnamefont {Z.}~\bibnamefont
  {Lu}}, \bibinfo {author} {\bibfnamefont {X.}~\bibnamefont {Wang}},\ and\
  \bibinfo {author} {\bibfnamefont {Y.}~\bibnamefont {Li}},\ }\bibinfo {title}
  {Experimental demonstration of continuous-variable
  measurement-device-independent quantum key distribution over optical fiber},\
  \href@noop {} {\bibfield  {journal} {\bibinfo  {journal} {Optica}\ }\textbf
  {\bibinfo {volume} {9}},\ \bibinfo {pages} {492} (\bibinfo {year}
  {2022})}\BibitemShut {NoStop}%
\bibitem [{\citenamefont {Furusawa}\ \emph
  {et~al.}(1998{\natexlab{a}})\citenamefont {Furusawa}, \citenamefont
  {S{\o}rensen}, \citenamefont {Braunstein}, \citenamefont {Fuchs},
  \citenamefont {Kimble},\ and\ \citenamefont
  {Polzik}}]{30BellM_furusawa1998unconditional}%
  \BibitemOpen
  \bibfield  {author} {\bibinfo {author} {\bibfnamefont {A.}~\bibnamefont
  {Furusawa}}, \bibinfo {author} {\bibfnamefont {J.~L.}\ \bibnamefont
  {S{\o}rensen}}, \bibinfo {author} {\bibfnamefont {S.~L.}\ \bibnamefont
  {Braunstein}}, \bibinfo {author} {\bibfnamefont {C.~A.}\ \bibnamefont
  {Fuchs}}, \bibinfo {author} {\bibfnamefont {H.~J.}\ \bibnamefont {Kimble}},\
  and\ \bibinfo {author} {\bibfnamefont {E.~S.}\ \bibnamefont {Polzik}},\
  }\bibinfo {title} {Unconditional quantum teleportation},\ \href@noop {}
  {\bibfield  {journal} {\bibinfo  {journal} {science}\ }\textbf {\bibinfo
  {volume} {282}},\ \bibinfo {pages} {706} (\bibinfo {year}
  {1998}{\natexlab{a}})}\BibitemShut {NoStop}%
\bibitem [{\citenamefont {Polkinghorne}\ and\ \citenamefont
  {Ralph}(1999{\natexlab{a}})}]{31BellM_polkinghorne1999continuous}%
  \BibitemOpen
  \bibfield  {author} {\bibinfo {author} {\bibfnamefont {R.}~\bibnamefont
  {Polkinghorne}}\ and\ \bibinfo {author} {\bibfnamefont {T.}~\bibnamefont
  {Ralph}},\ }\bibinfo {title} {Continuous variable entanglement swapping},\
  \href@noop {} {\bibfield  {journal} {\bibinfo  {journal} {Phys. Rev. Lett.}\
  }\textbf {\bibinfo {volume} {83}},\ \bibinfo {pages} {2095} (\bibinfo {year}
  {1999}{\natexlab{a}})}\BibitemShut {NoStop}%
\bibitem [{\citenamefont {Roumestan}\ \emph {et~al.}(2024)\citenamefont
  {Roumestan}, \citenamefont {Ghazisaeidi}, \citenamefont {Renaudier},
  \citenamefont {Vidarte}, \citenamefont {Leverrier}, \citenamefont
  {Diamanti},\ and\ \citenamefont
  {Grangier}}]{33highspeed_roumestan2024shaped}%
  \BibitemOpen
  \bibfield  {author} {\bibinfo {author} {\bibfnamefont {F.}~\bibnamefont
  {Roumestan}}, \bibinfo {author} {\bibfnamefont {A.}~\bibnamefont
  {Ghazisaeidi}}, \bibinfo {author} {\bibfnamefont {J.}~\bibnamefont
  {Renaudier}}, \bibinfo {author} {\bibfnamefont {L.~T.}\ \bibnamefont
  {Vidarte}}, \bibinfo {author} {\bibfnamefont {A.}~\bibnamefont {Leverrier}},
  \bibinfo {author} {\bibfnamefont {E.}~\bibnamefont {Diamanti}},\ and\
  \bibinfo {author} {\bibfnamefont {P.}~\bibnamefont {Grangier}},\ }\bibinfo
  {title} {Shaped constellation continuous variable quantum key distribution:
  Concepts, methods and experimental validation},\ \href@noop {} {\bibfield
  {journal} {\bibinfo  {journal} {J. Lightwave Technol.}\ } (\bibinfo {year}
  {2024})}\BibitemShut {NoStop}%
\bibitem [{\citenamefont {Hajomer}\ \emph
  {et~al.}(2024{\natexlab{d}})\citenamefont {Hajomer}, \citenamefont
  {Bruynsteen}, \citenamefont {Derkach}, \citenamefont {Jain}, \citenamefont
  {Bomhals}, \citenamefont {Bastiaens}, \citenamefont {Andersen}, \citenamefont
  {Yin},\ and\ \citenamefont {Gehring}}]{34highspeed_hajomer2024continuous}%
  \BibitemOpen
  \bibfield  {author} {\bibinfo {author} {\bibfnamefont {A.~A.}\ \bibnamefont
  {Hajomer}}, \bibinfo {author} {\bibfnamefont {C.}~\bibnamefont {Bruynsteen}},
  \bibinfo {author} {\bibfnamefont {I.}~\bibnamefont {Derkach}}, \bibinfo
  {author} {\bibfnamefont {N.}~\bibnamefont {Jain}}, \bibinfo {author}
  {\bibfnamefont {A.}~\bibnamefont {Bomhals}}, \bibinfo {author} {\bibfnamefont
  {S.}~\bibnamefont {Bastiaens}}, \bibinfo {author} {\bibfnamefont {U.~L.}\
  \bibnamefont {Andersen}}, \bibinfo {author} {\bibfnamefont {X.}~\bibnamefont
  {Yin}},\ and\ \bibinfo {author} {\bibfnamefont {T.}~\bibnamefont {Gehring}},\
  }\bibinfo {title} {Continuous-variable quantum key distribution at 10 gbaud
  using an integrated photonic-electronic receiver},\ \href@noop {} {\bibfield
  {journal} {\bibinfo  {journal} {Optica}\ }\textbf {\bibinfo {volume} {11}},\
  \bibinfo {pages} {1197} (\bibinfo {year} {2024}{\natexlab{d}})}\BibitemShut
  {NoStop}%
\bibitem [{\citenamefont {Wright}\ \emph {et~al.}(2017)\citenamefont {Wright},
  \citenamefont {Karpi{\'n}ski}, \citenamefont {S{\"o}ller},\ and\
  \citenamefont {Smith}}]{35wright2017spectral}%
  \BibitemOpen
  \bibfield  {author} {\bibinfo {author} {\bibfnamefont {L.~J.}\ \bibnamefont
  {Wright}}, \bibinfo {author} {\bibfnamefont {M.}~\bibnamefont
  {Karpi{\'n}ski}}, \bibinfo {author} {\bibfnamefont {C.}~\bibnamefont
  {S{\"o}ller}},\ and\ \bibinfo {author} {\bibfnamefont {B.~J.}\ \bibnamefont
  {Smith}},\ }\bibinfo {title} {Spectral shearing of quantum light pulses by
  electro-optic phase modulation},\ \href@noop {} {\bibfield  {journal}
  {\bibinfo  {journal} {Phys. Rev. Lett.}\ }\textbf {\bibinfo {volume} {118}},\
  \bibinfo {pages} {023601} (\bibinfo {year} {2017})}\BibitemShut {NoStop}%
\bibitem [{\citenamefont {Blow}\ \emph {et~al.}(1990)\citenamefont {Blow},
  \citenamefont {Loudon}, \citenamefont {Phoenix},\ and\ \citenamefont
  {Shepherd}}]{36TM_blow1990continuum}%
  \BibitemOpen
  \bibfield  {author} {\bibinfo {author} {\bibfnamefont {K.}~\bibnamefont
  {Blow}}, \bibinfo {author} {\bibfnamefont {R.}~\bibnamefont {Loudon}},
  \bibinfo {author} {\bibfnamefont {S.~J.}\ \bibnamefont {Phoenix}},\ and\
  \bibinfo {author} {\bibfnamefont {T.}~\bibnamefont {Shepherd}},\ }\bibinfo
  {title} {Continuum fields in quantum optics},\ \href@noop {} {\bibfield
  {journal} {\bibinfo  {journal} {Phys. Rev. A}\ }\textbf {\bibinfo {volume}
  {42}},\ \bibinfo {pages} {4102} (\bibinfo {year} {1990})}\BibitemShut
  {NoStop}%
\bibitem [{\citenamefont {Brecht}\ \emph {et~al.}(2015)\citenamefont {Brecht},
  \citenamefont {Reddy}, \citenamefont {Silberhorn},\ and\ \citenamefont
  {Raymer}}]{37TM_brecht2015photon}%
  \BibitemOpen
  \bibfield  {author} {\bibinfo {author} {\bibfnamefont {B.}~\bibnamefont
  {Brecht}}, \bibinfo {author} {\bibfnamefont {D.~V.}\ \bibnamefont {Reddy}},
  \bibinfo {author} {\bibfnamefont {C.}~\bibnamefont {Silberhorn}},\ and\
  \bibinfo {author} {\bibfnamefont {M.~G.}\ \bibnamefont {Raymer}},\ }\bibinfo
  {title} {Photon temporal modes: a complete framework for quantum information
  science},\ \href@noop {} {\bibfield  {journal} {\bibinfo  {journal} {Phys.
  Rev. X}\ }\textbf {\bibinfo {volume} {5}},\ \bibinfo {pages} {041017}
  (\bibinfo {year} {2015})}\BibitemShut {NoStop}%
\bibitem [{\citenamefont {Fabre}\ and\ \citenamefont
  {Treps}(2020)}]{38TM_fabre2020modes}%
  \BibitemOpen
  \bibfield  {author} {\bibinfo {author} {\bibfnamefont {C.}~\bibnamefont
  {Fabre}}\ and\ \bibinfo {author} {\bibfnamefont {N.}~\bibnamefont {Treps}},\
  }\bibinfo {title} {Modes and states in quantum optics},\ \href@noop {}
  {\bibfield  {journal} {\bibinfo  {journal} {Rev. Mod. Phys.}\ }\textbf
  {\bibinfo {volume} {92}},\ \bibinfo {pages} {035005} (\bibinfo {year}
  {2020})}\BibitemShut {NoStop}%
\bibitem [{\citenamefont {Raymer}\ and\ \citenamefont
  {Walmsley}(2020)}]{39TM_raymer2020temporal}%
  \BibitemOpen
  \bibfield  {author} {\bibinfo {author} {\bibfnamefont {M.~G.}\ \bibnamefont
  {Raymer}}\ and\ \bibinfo {author} {\bibfnamefont {I.~A.}\ \bibnamefont
  {Walmsley}},\ }\bibinfo {title} {Temporal modes in quantum optics: then and
  now},\ \href@noop {} {\bibfield  {journal} {\bibinfo  {journal} {Phys. Scr.}\
  }\textbf {\bibinfo {volume} {95}},\ \bibinfo {pages} {064002} (\bibinfo
  {year} {2020})}\BibitemShut {NoStop}%
\bibitem [{\citenamefont {Zhao}\ \emph {et~al.}(2021)\citenamefont {Zhao},
  \citenamefont {Huo}, \citenamefont {Cui}, \citenamefont {Li},\ and\
  \citenamefont {Ou}}]{40TM_zhao2021propagation}%
  \BibitemOpen
  \bibfield  {author} {\bibinfo {author} {\bibfnamefont {W.}~\bibnamefont
  {Zhao}}, \bibinfo {author} {\bibfnamefont {N.}~\bibnamefont {Huo}}, \bibinfo
  {author} {\bibfnamefont {L.}~\bibnamefont {Cui}}, \bibinfo {author}
  {\bibfnamefont {X.}~\bibnamefont {Li}},\ and\ \bibinfo {author}
  {\bibfnamefont {Z.}~\bibnamefont {Ou}},\ }\bibinfo {title} {Propagation of
  temporal mode multiplexed optical fields in fibers: influence of
  dispersion},\ \href@noop {} {\bibfield  {journal} {\bibinfo  {journal} {Opt.
  Express}\ }\textbf {\bibinfo {volume} {30}},\ \bibinfo {pages} {447}
  (\bibinfo {year} {2021})}\BibitemShut {NoStop}%
\bibitem [{\citenamefont {Raymer}\ \emph {et~al.}(1989)\citenamefont {Raymer},
  \citenamefont {Li},\ and\ \citenamefont
  {Walmsley}}]{41TM_raymer1989temporal}%
  \BibitemOpen
  \bibfield  {author} {\bibinfo {author} {\bibfnamefont {M.}~\bibnamefont
  {Raymer}}, \bibinfo {author} {\bibfnamefont {Z.}~\bibnamefont {Li}},\ and\
  \bibinfo {author} {\bibfnamefont {I.}~\bibnamefont {Walmsley}},\ }\bibinfo
  {title} {Temporal quantum fluctuations in stimulated raman scattering:
  Coherent-modes description},\ \href@noop {} {\bibfield  {journal} {\bibinfo
  {journal} {Phys. Rev. Lett.}\ }\textbf {\bibinfo {volume} {63}},\ \bibinfo
  {pages} {1586} (\bibinfo {year} {1989})}\BibitemShut {NoStop}%
\bibitem [{\citenamefont {Loudon}(2000)}]{42loudon2000quantum}%
  \BibitemOpen
  \bibfield  {author} {\bibinfo {author} {\bibfnamefont {R.}~\bibnamefont
  {Loudon}},\ }\bibinfo {title} {The quantum theory of light},\ \href@noop {}
  {\emph {\bibinfo {title} {The quantum theory of light}}}\ (\bibinfo
  {publisher} {OUP Oxford},\ \bibinfo {year} {2000})\BibitemShut {NoStop}%
\bibitem [{\citenamefont {Grosshans}\ \emph {et~al.}(2003)\citenamefont
  {Grosshans}, \citenamefont {Cerf}, \citenamefont {Wenger}, \citenamefont
  {Tualle-Brouri},\ and\ \citenamefont {Grangier}}]{grosshans2003virtual}%
  \BibitemOpen
  \bibfield  {author} {\bibinfo {author} {\bibfnamefont {F.}~\bibnamefont
  {Grosshans}}, \bibinfo {author} {\bibfnamefont {N.~J.}\ \bibnamefont {Cerf}},
  \bibinfo {author} {\bibfnamefont {J.}~\bibnamefont {Wenger}}, \bibinfo
  {author} {\bibfnamefont {R.}~\bibnamefont {Tualle-Brouri}},\ and\ \bibinfo
  {author} {\bibfnamefont {P.}~\bibnamefont {Grangier}},\ }\bibinfo {title}
  {Virtual entanglement and reconciliation protocols for quantum cryptography
  with continuous variables},\ \href@noop {} {\bibfield  {journal} {\bibinfo
  {journal} {arXiv preprint quant-ph/0306141}\ } (\bibinfo {year}
  {2003})}\BibitemShut {NoStop}%
\bibitem [{\citenamefont {Furusawa}\ \emph
  {et~al.}(1998{\natexlab{b}})\citenamefont {Furusawa}, \citenamefont
  {S{\o}rensen}, \citenamefont {Braunstein}, \citenamefont {Fuchs},
  \citenamefont {Kimble},\ and\ \citenamefont
  {Polzik}}]{furusawa1998unconditional}%
  \BibitemOpen
  \bibfield  {author} {\bibinfo {author} {\bibfnamefont {A.}~\bibnamefont
  {Furusawa}}, \bibinfo {author} {\bibfnamefont {J.~L.}\ \bibnamefont
  {S{\o}rensen}}, \bibinfo {author} {\bibfnamefont {S.~L.}\ \bibnamefont
  {Braunstein}}, \bibinfo {author} {\bibfnamefont {C.~A.}\ \bibnamefont
  {Fuchs}}, \bibinfo {author} {\bibfnamefont {H.~J.}\ \bibnamefont {Kimble}},\
  and\ \bibinfo {author} {\bibfnamefont {E.~S.}\ \bibnamefont {Polzik}},\
  }\bibinfo {title} {Unconditional quantum teleportation},\ \href@noop {}
  {\bibfield  {journal} {\bibinfo  {journal} {Science}\ }\textbf {\bibinfo
  {volume} {282}},\ \bibinfo {pages} {706} (\bibinfo {year}
  {1998}{\natexlab{b}})}\BibitemShut {NoStop}%
\bibitem [{\citenamefont {Polkinghorne}\ and\ \citenamefont
  {Ralph}(1999{\natexlab{b}})}]{polkinghorne1999continuous}%
  \BibitemOpen
  \bibfield  {author} {\bibinfo {author} {\bibfnamefont {R.}~\bibnamefont
  {Polkinghorne}}\ and\ \bibinfo {author} {\bibfnamefont {T.}~\bibnamefont
  {Ralph}},\ }\bibinfo {title} {Continuous variable entanglement swapping},\
  \href@noop {} {\bibfield  {journal} {\bibinfo  {journal} {Phys. Rev. Lett.}\
  }\textbf {\bibinfo {volume} {83}},\ \bibinfo {pages} {2095} (\bibinfo {year}
  {1999}{\natexlab{b}})}\BibitemShut {NoStop}%
\bibitem [{\citenamefont {Cover}(1999)}]{43Eq4_cover1999elements}%
  \BibitemOpen
  \bibfield  {author} {\bibinfo {author} {\bibfnamefont {T.~M.}\ \bibnamefont
  {Cover}},\ }\bibinfo {title} {Elements of information theory},\ \href@noop {}
  {\emph {\bibinfo {title} {Elements of information theory}}}\ (\bibinfo
  {publisher} {John Wiley \& Sons},\ \bibinfo {year} {1999})\BibitemShut
  {NoStop}%
\bibitem [{\citenamefont {Devetak}\ and\ \citenamefont
  {Winter}(2005)}]{devetak2005distillation}%
  \BibitemOpen
  \bibfield  {author} {\bibinfo {author} {\bibfnamefont {I.}~\bibnamefont
  {Devetak}}\ and\ \bibinfo {author} {\bibfnamefont {A.}~\bibnamefont
  {Winter}},\ }\bibinfo {title} {Distillation of secret key and entanglement
  from quantum states},\ \href@noop {} {\bibfield  {journal} {\bibinfo
  {journal} {Proceedings of the Royal Society A: Mathematical, Physical and
  engineering sciences}\ }\textbf {\bibinfo {volume} {461}},\ \bibinfo {pages}
  {207} (\bibinfo {year} {2005})}\BibitemShut {NoStop}%
\bibitem [{\citenamefont {Leverrier}\ \emph {et~al.}(2010)\citenamefont
  {Leverrier}, \citenamefont {Grosshans},\ and\ \citenamefont
  {Grangier}}]{45finitesize_leverrier2010finite}%
  \BibitemOpen
  \bibfield  {author} {\bibinfo {author} {\bibfnamefont {A.}~\bibnamefont
  {Leverrier}}, \bibinfo {author} {\bibfnamefont {F.}~\bibnamefont
  {Grosshans}},\ and\ \bibinfo {author} {\bibfnamefont {P.}~\bibnamefont
  {Grangier}},\ }\bibinfo {title} {Finite-size analysis of a
  continuous-variable quantum key distribution},\ \href@noop {} {\bibfield
  {journal} {\bibinfo  {journal} {Phys. Rev. A}\ }\textbf {\bibinfo {volume}
  {81}},\ \bibinfo {pages} {062343} (\bibinfo {year} {2010})}\BibitemShut
  {NoStop}%
\bibitem [{\citenamefont {Furrer}\ \emph {et~al.}(2012)\citenamefont {Furrer},
  \citenamefont {Franz}, \citenamefont {Berta}, \citenamefont {Leverrier},
  \citenamefont {Scholz}, \citenamefont {Tomamichel},\ and\ \citenamefont
  {Werner}}]{46finitesize_furrer2012continuous}%
  \BibitemOpen
  \bibfield  {author} {\bibinfo {author} {\bibfnamefont {F.}~\bibnamefont
  {Furrer}}, \bibinfo {author} {\bibfnamefont {T.}~\bibnamefont {Franz}},
  \bibinfo {author} {\bibfnamefont {M.}~\bibnamefont {Berta}}, \bibinfo
  {author} {\bibfnamefont {A.}~\bibnamefont {Leverrier}}, \bibinfo {author}
  {\bibfnamefont {V.~B.}\ \bibnamefont {Scholz}}, \bibinfo {author}
  {\bibfnamefont {M.}~\bibnamefont {Tomamichel}},\ and\ \bibinfo {author}
  {\bibfnamefont {R.~F.}\ \bibnamefont {Werner}},\ }\bibinfo {title}
  {Continuous variable quantum key distribution: finite-key analysis of
  composable security against coherent attacks},\ \href@noop {} {\bibfield
  {journal} {\bibinfo  {journal} {Phys. Rev. Lett.}\ }\textbf {\bibinfo
  {volume} {109}},\ \bibinfo {pages} {100502} (\bibinfo {year}
  {2012})}\BibitemShut {NoStop}%
\bibitem [{\citenamefont {Zhang}\ \emph {et~al.}(2017)\citenamefont {Zhang},
  \citenamefont {Zhang}, \citenamefont {Zhao}, \citenamefont {Wang},
  \citenamefont {Yu},\ and\ \citenamefont {Guo}}]{zhang2017finite}%
  \BibitemOpen
  \bibfield  {author} {\bibinfo {author} {\bibfnamefont {X.}~\bibnamefont
  {Zhang}}, \bibinfo {author} {\bibfnamefont {Y.}~\bibnamefont {Zhang}},
  \bibinfo {author} {\bibfnamefont {Y.}~\bibnamefont {Zhao}}, \bibinfo {author}
  {\bibfnamefont {X.}~\bibnamefont {Wang}}, \bibinfo {author} {\bibfnamefont
  {S.}~\bibnamefont {Yu}},\ and\ \bibinfo {author} {\bibfnamefont
  {H.}~\bibnamefont {Guo}},\ }\bibinfo {title} {Finite-size analysis of
  continuous-variable measurement-device-independent quantum key
  distribution},\ \href@noop {} {\bibfield  {journal} {\bibinfo  {journal}
  {Phys. Rev. A}\ }\textbf {\bibinfo {volume} {96}},\ \bibinfo {pages} {042334}
  (\bibinfo {year} {2017})}\BibitemShut {NoStop}%
\bibitem [{\citenamefont {Lupo}\ \emph
  {et~al.}(2018{\natexlab{b}})\citenamefont {Lupo}, \citenamefont {Ottaviani},
  \citenamefont {Papanastasiou},\ and\ \citenamefont
  {Pirandola}}]{lupo2018continuousYORK}%
  \BibitemOpen
  \bibfield  {author} {\bibinfo {author} {\bibfnamefont {C.}~\bibnamefont
  {Lupo}}, \bibinfo {author} {\bibfnamefont {C.}~\bibnamefont {Ottaviani}},
  \bibinfo {author} {\bibfnamefont {P.}~\bibnamefont {Papanastasiou}},\ and\
  \bibinfo {author} {\bibfnamefont {S.}~\bibnamefont {Pirandola}},\ }\bibinfo
  {title} {Continuous-variable measurement-device-independent quantum key
  distribution: Composable security against coherent attacks},\ \href@noop {}
  {\bibfield  {journal} {\bibinfo  {journal} {Phys. Rev. A}\ }\textbf {\bibinfo
  {volume} {97}},\ \bibinfo {pages} {052327} (\bibinfo {year}
  {2018}{\natexlab{b}})}\BibitemShut {NoStop}%
\end{thebibliography}%
\end{document}